\documentclass[
aps,
reprint,
superscriptaddress,
amsmath,amssymb,
floatfix,
]{revtex4-2}
\usepackage[utf8]{inputenc}
\usepackage[T1]{fontenc}
\usepackage{color}
\usepackage{microtype}
\usepackage{graphicx}
\usepackage{dcolumn}
\usepackage{bm}
\usepackage{gensymb}
\usepackage{orcidlink}
\usepackage{physics}
\usepackage{xfrac}
\usepackage{amsmath}
\usepackage{multirow}
\usepackage{natbib}

\begin{document}

\title{Nuclear charge radii of aluminium isotopes at the proton drip line}

\author{A.~J.~Brinson\,\orcidlink{0000-0002-9551-5298}}
\email{brinson@mit.edu}
\affiliation{Laboratory for Nuclear Science, Massachusetts Institute of Technology, Cambridge, Massachusetts 02139, USA}

\author{B.~J.~Rickey\,\orcidlink{0009-0004-5935-4505}}
\email{rickey@tamu.edu}
\altaffiliation{Present Address: The Cyclotron Institute at Texas A\&M University, College Station, Texas 77840, USA}
\affiliation{Facility for Rare Isotope Beams, Michigan State University, East Lansing, Michigan 48824, USA}
\affiliation{Department of Physics and Astronomy, Michigan State University, East Lansing, Michigan 48824, USA}

\author{P.~Arthuis\,\orcidlink{0000-0002-7073-9340}}
\affiliation{Université Paris-Saclay, CNRS/IN2P3, IJCLab, 91405 Orsay, France}

\author{A.~Belley\,\orcidlink{0000-0002-0088-9714}}
\affiliation{Laboratory for Nuclear Science, Massachusetts Institute of Technology, Cambridge, Massachusetts 02139, USA}

\author{S.~E.~Campbell\,\orcidlink{0009-0005-2053-2711}}
\affiliation{Facility for Rare Isotope Beams, Michigan State University, East Lansing, Michigan 48824, USA}
\affiliation{Department of Physics and Astronomy, Michigan State University, East Lansing, Michigan 48824, USA}

\author{X.~Chen\,\orcidlink{0000-0003-3513-8870}}
\affiliation{Facility for Rare Isotope Beams, Michigan State University, East Lansing, Michigan 48824, USA}

\author{A.~Dockery\,\orcidlink{0000-0002-3066-978X}}
\affiliation{Facility for Rare Isotope Beams, Michigan State University, East Lansing, Michigan 48824, USA}
\affiliation{Department of Physics and Astronomy, Michigan State University, East Lansing, Michigan 48824, USA}

\author{S.~Elhatisari\,\orcidlink{0000-0002-7951-1991}}
\affiliation{Faculty of Natural Sciences and Engineering, Gaziantep Islam Science and Technology University, Gaziantep 27010, Turkey}

\author{H.~Erington,\,\orcidlink{0009-0005-4433-3020}}
\affiliation{Facility for Rare Isotope Beams, Michigan State University, East Lansing, Michigan 48824, USA}
\affiliation{Department of Physics and Astronomy, Michigan State University, East Lansing, Michigan 48824, USA}

\author{N.~D.~Gamage,\,\orcidlink{0000-0001-5657-2081}}
\affiliation{Facility for Rare Isotope Beams, Michigan State University, East Lansing, Michigan 48824, USA}

\author{R.~F.~Garcia~Ruiz\,\orcidlink{0000-0002-2926-5569}}
\email{rgarciar@mit.edu}
\affiliation{Laboratory for Nuclear Science, Massachusetts Institute of Technology, Cambridge, Massachusetts 02139, USA}
\affiliation{Harvard-MIT Center for Ultracold Atoms, Cambridge, Massachusetts 02138, USA}

\author{M.~Heinz\,\orcidlink{0000-0002-6363-0056}}
\affiliation{National Center for Computational Sciences, Oak Ridge National Laboratory, Oak Ridge, TN 37831, USA}
\affiliation{Physics Division, Oak Ridge National Laboratory, Oak Ridge, TN 37831, USA}

\author{J.~D.~ Holt}
\affiliation{TRIUMF, Vancouver, BC V6T 2A3, Canada}
\affiliation{Department of Physics, McGill University, Montr\'eal, QC H3A 2T8, Canada}

\author{C.~M.~Ireland\,\orcidlink{0009-0002-4079-1567}}
\affiliation{Facility for Rare Isotope Beams, Michigan State University, East Lansing, Michigan 48824, USA}
\affiliation{Department of Physics and Astronomy, Michigan State University, East Lansing, Michigan 48824, USA}

\author{C.~Izzo\,\orcidlink{0000-0002-4119-1654}}
\affiliation{Facility for Rare Isotope Beams, Michigan State University, East Lansing, Michigan 48824, USA}

\author{C.~M.~Jones}
\affiliation{Department of Biological and Physical Sciences, South Carolina State University, Orangeburg, South Carolina, USA}

\author{J.~Karthein\,\orcidlink{0000-0002-4306-9708}}
\altaffiliation{Present Address: Cyclotron Institute and Department of Physics \& Astronomy, Texas A\&M University, College Station, Texas 77840, USA}
\affiliation{Laboratory for Nuclear Science, Massachusetts Institute of Technology, Cambridge, Massachusetts 02139, USA}

\author{K.~König\,\orcidlink{0000-0001-9415-3208}}
\affiliation{Institut f\"ur Kernphysik, Technische Universit\"at Darmstadt, 64289 Darmstadt, Germany}%
\affiliation{Helmholtz Research Academy Hesse for FAIR, GSI Darmstadt, 64291 Darmstadt, Germany}

\author{D.~J.~Lee\,\orcidlink{0000-0002-3630-567X}}
\affiliation{Facility for Rare Isotope Beams, Michigan State University, East Lansing, Michigan 48824, USA}
\affiliation{Department of Physics and Astronomy, Michigan State University, East Lansing, Michigan 48824, USA}

\author{Y.-Z.~Ma\,\orcidlink{0000-0002-0892-4457}}
\affiliation{Facility for Rare Isotope Beams, Michigan State University, East Lansing, Michigan 48824, USA}
\affiliation{Department of Physics and Astronomy, Michigan State University, East Lansing, Michigan 48824, USA}

\author{F.~M.~Maier\,\orcidlink{0000-0001-9715-2147}}
\affiliation{Facility for Rare Isotope Beams, Michigan State University, East Lansing, Michigan 48824, USA}

\author{U.-G.~Mei{\ss}ner\,\orcidlink{0000-0003-1254-442X}}
\affiliation{Helmholtz-Institut fur Strahlen- und Kernphysik and Bethe Center for Theoretical Physics, Universit\"at Bonn, D-53115 Bonn, Germany}
\affiliation{Institute for Advanced Simulation (IAS-4), Forschungszentrum J\"ulich, D-52425 J\"ulich, Germany}

\author{K.~Minamisono\,\orcidlink{0000-0003-2315-5032} }
\email{minamiso@frib.msu.edu}
\affiliation{Facility for Rare Isotope Beams, Michigan State University, East Lansing, Michigan 48824, USA}
\affiliation{Department of Physics and Astronomy, Michigan State University, East Lansing, Michigan 48824, USA}

\author{M.~D.~Moenter\,\orcidlink{0009-0004-0177-2210}}
\altaffiliation{Present Address: Cyclotron Institute and Department of Physics \& Astronomy, Texas A\&M University, College Station, Texas 77840, USA}
\affiliation{Facility for Rare Isotope Beams, Michigan State University, East Lansing, Michigan 48824, USA}
\affiliation{Department of Physics and Astronomy, Michigan State University, East Lansing, Michigan 48824, USA}

\author{J.~M.~Munoz\,\orcidlink{0000-0001-7740-2866}}
\affiliation{Laboratory for Nuclear Science, Massachusetts Institute of Technology, Cambridge, Massachusetts 02139, USA}

\author{W.~N\"ortersh\"auser\,\orcidlink{0000-0001-7432-3687}}
\affiliation{Institut f\"ur Kernphysik, Technische Universit\"at Darmstadt, 64289 Darmstadt, Germany}%
\affiliation{Helmholtz Research Academy Hesse for FAIR, GSI Darmstadt, 64291 Darmstadt, Germany}

\author{A.~Ortiz-Cortes\,\orcidlink{0009-0000-3491-2006}}
\affiliation{Facility for Rare Isotope Beams, Michigan State University, East Lansing, Michigan 48824, USA}

\author{J.~Palmes\,\orcidlink{0009-0007-3928-7095}}
\affiliation{Institut f\"ur Kernphysik, Technische Universit\"at Darmstadt, 64289 Darmstadt, Germany}%

\author{S.~Papa}
\affiliation{Department of Biological and Physical Sciences, South Carolina State University, Orangeburg, South Carolina, USA}

\author{F.~C.~Pastrana~Cruz\,\orcidlink{0009-0008-7469-7513}}
\affiliation{Laboratory for Nuclear Science, Massachusetts Institute of Technology, Cambridge, Massachusetts 02139, USA}

\author{R.~Ringle}
\affiliation{Facility for Rare Isotope Beams, Michigan State University, East Lansing, Michigan 48824, USA}
\affiliation{Department of Physics and Astronomy, Michigan State University, East Lansing, Michigan 48824, USA}

\author{H.~Sims}
\affiliation{Facility for Rare Isotope Beams, Michigan State University, East Lansing, Michigan 48824, USA}
\affiliation{Department of Physics and Astronomy, Michigan State University, East Lansing, Michigan 48824, USA}

\author{C.~Sumithrarachchi }
\affiliation{Facility for Rare Isotope Beams, Michigan State University, East Lansing, Michigan 48824, USA}

\author{A.~R.~Vernon\,\orcidlink{0000-0001-8130-0109}}
\affiliation{Laboratory for Nuclear Science, Massachusetts Institute of Technology, Cambridge, Massachusetts 02139, USA}
\affiliation{Department of Physics, William \& Mary, 300 Ukrop Way, Williamsburg, VA 23185, USA}

\author{T.~Wang\,\orcidlink{0000-0002-0457-3135}}
\affiliation{School of Physics, Peking University, Beijing 100871, China}

\author{S.~G.~Wilkins\,\orcidlink{0000-0001-8897-7227}}
\altaffiliation{Present Address: Facility for Rare Isotope Beams, Michigan State University, East Lansing, USA}
\affiliation{Laboratory for Nuclear Science, Massachusetts Institute of Technology, Cambridge, Massachusetts 02139, USA}

\author{R.~B.~Yadav\,\orcidlink{0000-0003-2801-823X}}
\affiliation{Department of Biological and Physical Sciences, South Carolina State University, Orangeburg, South Carolina, USA}

\author{S.~Zhang\,\orcidlink{0000-0003-0811-3432}}
\affiliation{Institute for Advanced Simulation (IAS-4), Forschungszentrum J\"ulich, D-52425 J\"ulich, Germany}

\date{\today}

\begin{abstract}
Understanding the evolution of nuclear size away from stability remains a central challenge in nuclear physics. 
In neutron-deficient systems, charge radii can be highly sensitive to the interplay between strong and electromagnetic interactions, and the effects of weak binding, giving rise to exotic nuclear phenomena.
However, experimental data on these systems has been limited by short lifetimes and low production rates. 
Here we report the first laser-spectroscopy measurements of nuclear charge radii along the neutron-deficient aluminium isotopic chain, from $^{25}$Al to the proton-drip-line nucleus $^{22}$Al, using the {Resonance Ionization Spectroscopy Experiment} (RISE) at the {Facility for Rare Isotope Beams} (FRIB). 
Our measurements reveal a step-like increase in charge radius toward the drip line, with similar radii for $^{22,\,23}$Al. A comparison of our results with those of their mirror partners reveals an almost identical correlation with the calculated proton skins and is consistent with the systematic trend of well-bound nuclei.
These results offer insight for understanding the evolution of nuclear size at the proton dripline and place important constraints on modern nuclear theory. They also demonstrate the unique combined capabilities of RISE and FRIB to probe the structures of previously inaccessible nuclei at the limits of existence.
\end{abstract}

\maketitle


The size of the atomic nucleus encodes information about the details of the interactions among its constituent protons and neutrons. Experimental knowledge of nuclear radii can therefore provide critical guidance for the development of nuclear theory.
Over the past decades, high-resolution laser spectroscopy has made major contributions to this field \cite{Yan22}, revealing subtle trends in charge radii across long isotopic chains and exposing unexpected structural features such as rapid changes of the nuclear size in neutron-rich nuclei \cite{K-radii2021,Ca-radii2016,Karthein2024}, odd–even staggering \cite{Cu-radii2020}, the continuum effect \cite{mil19,kon23} and the emergence of halos in weakly bound systems \cite{Gei08,Lu13}. 

Several neutron-halo nuclei are now known near the neutron drip line \cite{tanihata1985measurements,Lu13}, whereas proton halos are much rarer because the Coulomb barrier suppresses extended proton distributions. For nuclei near stability, the charge distribution is governed primarily by the strong nuclear force, whereas
for neutron-deficient systems, the charge radius can be dramatically influenced by the balance between electromagnetic and strong interactions. 
Near the proton drip line, weak binding can lead to pronounced spatial extensions of nuclear densities \cite{Lin2026}.  
Typically, short lifetimes (often below one second) and low production yields have prevented direct measurements of nuclear charge radii for these nuclei.
In particular, several light neutron-deficient nuclei with proton number $Z<18$, whose properties can be described with high accuracy by modern \textit{ab initio} nuclear-structure calculations, have remained experimentally inaccessible to laser-spectroscopy studies, the technique of choice for extracting changes in the nuclear charge radii away from stability~\cite{Yan22}.

The aluminium isotopic chain ($Z = 13$) provides a particularly attractive testing ground for understanding the evolution of nuclear charge radii \cite{Hey21}. 
With a single proton hole below the sub-shell closure at $Z = 14$, the aluminium isotopes exhibit a rich interplay between single-particle and collective degrees of freedom \cite{Phi25}. 
On the neutron-deficient side, theory predicts unusual structural behavior driven by weak binding and deformation, and a possible proton-halo configuration has been suggested for $^{22\!}$Al~\cite{Lee2020}, as well as $^{23\!}$Al \cite{Cai2002}. 
Recent measurements have confirmed that $^{22\!}$Al is the last bound aluminium isotope \cite{Kostyleva2024Observation21}, with a proton-separation energy of only 100.4(8)~keV~\cite{Campbell2024, Sun2024Ground-stateCalculations}. However, a spin assignment of 4$^{+}$ disfavors a halo configuration in this nucleus \cite{Jensen2026_22Al_halo}.
Yet, until now, no experimental information on its charge radius, or those of its isotopic neighbors, was available. Here, we report the first laser-spectroscopy measurements of nuclear charge radii along the neutron-deficient aluminium isotopic chain, from \textsuperscript{25\!}Al down to the proton-drip-line nucleus \textsuperscript{22\!}Al. 
The measurements were performed using the newly commissioned {Resonance Ionization Spectroscopy Experiment (RISE)} \cite{Brinson2026RISE}, marking the first on-line operation of this instrument.

\section{The Resonance Ionization Spectroscopy Experiment (RISE)}
\begin{figure*}[]
\includegraphics[scale=0.7]{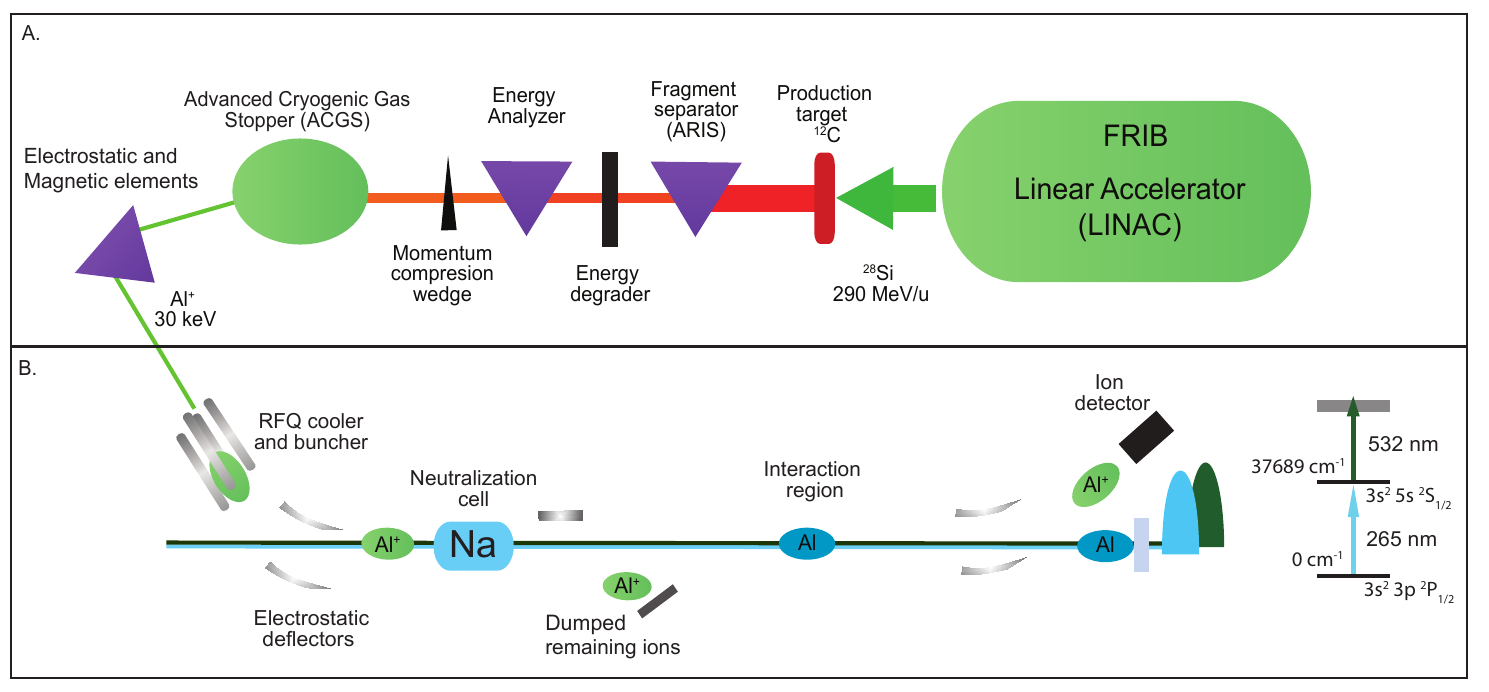} 
\caption{\label{fig:layout} \textbf{Experiment Layout.}
As shown in Panel A, a $^{28}$Si beam was accelerated to 290 MeV/u and impinged on a $^{12}$C target. The fragmentation products were then filtered by the Advanced Rare Isotope Separator (ARIS)~\cite{Hausmann2013} to isolate a desired isobar. The highly energetic Al isotopes were then decelerated and momentum compressed by an energy degrader and wedge, respectively, before being injected to the advanced cryogenic gas stopper (ACGS)~\cite{Lund2020OnlineNSCL}, where they were stopped, cooled, and subsequently redirected to the BECOLA facility in the stopped-beam experimental area in FRIB, shown in panel B. After the gas stopper, the ions were re-accelerated to $\sim$30~keV, mass separated with a magnetic dipole to remove stable contaminants resulting from the gas stopping process, and injected into a radio-frequency quadrupole cooler-buncher (RFQCB) \cite{Barquest2017RFQIsotopes} in the BECOLA facility. In the RFQCB, Al ions were collisionally cooled by a room-temperature He buffer gas and then stored for up to {10} ms before being released in bunches at a 100 Hz repetition rate synchronized with the laser pulses. The bunches were then accelerated to an energy of 30 keV \cite{kon24b} and passed through a Charge-Exchange Cell (CEC) \cite{Klose2012TestsSpectroscopy} for in-flight neutralization. The CEC was held at a variable potential with respect to the rest of the beamline to tune the bunch velocity prior to neutralization. Any remaining ions were removed from the bunches by an electrostatic ion deflector located directly after the CEC. The neutral bunch was then overlapped in succession by two laser pulses: (1) a 1~$\mu$J~265~nm pulse for resonant excitation, and (2) a 34~mJ 532~nm pulse for selective ionization. Atoms re-ionized by this laser sequence were steered onto a single-ion detector and correlated with the CEC potential and laser frequency to construct a hyperfine spectrum in their rest frame.}
\end{figure*}

RISE \cite{Brinson2026RISE} is a newly developed high-sensitivity instrument for collinear laser spectroscopy at the Facility for Rare Isotope Beams (FRIB). 
Installed as an extension to the BEam COoling LAser spectroscopy (BECOLA) facility \cite{Minamisono2013CommissioningNSCL,ros14} at FRIB, RISE combines a bunched low-energy ion beam with narrowband pulsed lasers to perform collinear resonance ionization spectroscopy~\cite{Sch91, Fla13,Cu-radii2020,K-radii2021} of short-lived isotopes with yields of only a few hundred ions per second. A layout of the RISE experiment integrated at FRIB is shown in Fig.~\ref{fig:layout}. 
Low-energy radioactive ions from the Advanced Cryogenic Gas Stopper (ACGS) \cite{Lund2020OnlineNSCL} are transported to the radio-frequency quadrupole cooler–buncher (RFQCB) \cite{Barquest2017RFQIsotopes} at the BECOLA facility. The singly charged ions are then cooled, bunched and accelerated to $\sim$30~keV and injected into the collinear laser spectroscopy apparatus. Prior to laser interaction, the ions are neutralized in flight using a sodium-vapor charge-exchange cell (CEC)\cite{Klose2012TestsSpectroscopy}. The resulting neutral atomic bunch is overlapped with multiple laser beams in an ultra-high-vacuum interaction region (pressure $\lesssim 5\times10^{-9}$~mbar), enabling collinear resonance ionization spectroscopy.
Resonant excitation is driven by a pulsed frequency-locked narrow-linewidth Ti:Sa laser, while a pulsed Nd:YAG laser is used for non-resonant ionization from the excited electronic state. The reionized atoms are subsequently deflected through a 45$\degree$ electrostatic bender and directed onto a MagneTOF ion detector. Hyperfine-structure spectra and their isotope shifts are obtained by varying the velocity of the initial ion bunch by applying a scanning voltage to the CEC, which is equivalent to Doppler tuning the frequency of the first-step laser in the laboratory frame.
Further details of the experimental setup can be found in Methods, as well as in Refs.~\cite{Brinson2026RISE,rickey2025halo,Rickey2026}.

The neutron-deficient aluminium isotopes were produced by fragmentation of a $^{28}$Si primary beam accelerated by the FRIB linear accelerator. The isotopes of interest were filtered by the ARIS fragment separator \cite{Portillo2023CommissioningFRIB} and then sent to the gas stopping cell. The stopped beam was then transported to the BECOLA RFQCB, where it was cooled via collisions with room temperature helium buffer gas. The cooled and bunched ions, with about 1~$\mu$s temporal width, were then accelerated to {29916.8(3)~eV}. The acceleration voltage was monitored by a high-voltage resistive divider and stabilized with a feedback system consisting of the voltage divider and a small power supply \cite{kon24b}.

A two-step resonant excitation and selective ionization scheme was used to measure the resonant structures of the $3s^23p\,\,^2\!P_{1/2} \rightarrow 3s^25s\,\,^2\!S_{1/2}$ transition in the neutral Al. For the resonant step, $\sim$1 $\mu$J pulses of~265~nm light were produced by locking an injection seeded cavity to a continuous-wave TiSa laser and single-pass tripling with a pair of BBO and BiBO crystals. The Matisse was locked at a different frequency for each isotope measured so that the anticollinear Doppler-shifted resonance would occur at the same nominal beam energy of 29916.8(3)~eV.
For selective ionization, 34 mJ pulses of nonresonant 532~nm light were produced by a frequency doubled Merion Nd:YAG laser. The nonresonant pulses were spatially and temporally overlapped with the excitation pulses, with a delay of 40~ns from the resonant laser pulse, to minimize the loss of resonant signal due to spontaneous decay of the excited state.

\section{Results and discussion}
Hyperfine spectra were collected for the short-lived isotopes $^{22,23,24,25\!}$Al, shown in Fig.~\ref{fig:iso2d}. For the production of $^{24\!}$Al, both the ground ($I^{\pi}=4^+$) and low-lying-isomeric ($I^{\pi}=1^+$) states were populated and transported to BECOLA. These states could not be isolated, and their hyperfine spectra were measured simultaneously. 
The recorded spectra were analyzed using a model that accounts for both the hyperfine splitting and the line-shape asymmetries arising from charge-exchange processes in the sodium vapor \cite{Dockery2025SpectralExchange}. Each spectrum was fitted with a Voigt profile to account for the various sources of line broadening. Details of the fitting procedure and data analysis are explained in Methods, `Measured spectra and data analysis'.

During the fitting, all hyperfine components were constrained to share common width and side-peak parameters. The amplitudes of the individual hyperfine components were treated as free parameters to allow independent adjustment of transition intensities.  For each isotope, $A$, the centroid frequency, $\nu_0^{A}$, was extracted from the best-fit parameters, and the centroid of $^{27\!}$Al was subtracted to calculate the isotope shift, $\delta\nu^{A,27}\equiv\nu_0^A-\nu_0^{27}$. The resulting isotope shifts for all measured isotopes are shown in Table~\ref{tab:Al_Shifts_Radii}. 

\begin{table*}[]
\caption{\textbf{Isotope shifts and extracted nuclear charge radii of neutron-deficient aluminium isotopes.} Isotope shifts, $\delta\nu^{A,27} \equiv \nu_0^{A}-\nu_0^{27}$, measured relative to $^{27\!}$Al. 
These isotope-shift values were used 
to extract the changes in the differential mean-square nuclear charge radii, 
$\delta\langle r^{2}\rangle_{\mathrm{ch}}$, using the field- and mass-shift 
constants, $F=70.11(13)$\,MHz/fm$^2$ and $K=-0.7(21)$\,GHz$\cdot$u \cite{Skripnikov2024Isotope-shift26m}, respectively. Absolute nuclear charge radii were obtained with reference to the charge radius of $^{27\!}$Al, 
$R_{\mathrm{ch}}(^{27\!}\mathrm{Al}) = 3.0614(29)\,\mathrm{fm}$, which is based on a recent reevaluation \cite{Ohayon2025} and a new calculation of nuclear polarization effects. Statistical, systematic, and atomic theory uncertainties are given in parentheses, square brackets, and curly braces, respectively.
The isomer shift between the isomeric and the ground state of $^{24\!}$Al can be obtained with higher accuracy, since both states are recorded simultaneously in a single spectrum and the isomer shift has a negligible mass-shift contribution. The results are $\delta\nu^{24\mathrm{m},24}=6.6(29)\rm{MHz}$ and $\delta\langle r^2_\mathrm{ch}\rangle^{24\mathrm{m}, 24}=0.0940(410)\{2\}\rm{fm}^2$, which corresponds to an isomeric radius difference of $\delta R_{\mathrm{ch}}=0.015(6)$\,fm. The influence of a possible hyperfine structure anomaly on the isotope and isomer shifts is discussed in Methods, `Systematic uncertainties' and is expected to be considerably smaller than the statistical uncertainty of our results. \\
}
\begin{tabular}{r|r|r|r|r}
\label{tab:Al_Shifts_Radii}
Isotope & I$^\pi$ &$\delta\nu^{A, 27}$ (stat)[sys] & $\delta\langle r^2_\mathrm{ch}\rangle^{A, 27}$ (stat)[sys]\{th\} & $\sqrt{\langle r^2_\mathrm{ch}}\rangle$ (tot) \\
\\[-2.5ex] \hline & \\[-2.5ex]
\rule{0mm}{3mm}\textsuperscript{25\!}Al & 5/2$^+$& $-2.5(17)$[39]\,MHz &$-0.01$(2)[6]\{ 9\}\,fm$^2$ & 3.060(18)\,fm \\
\textsuperscript{24\!}Al &   4$^+$& $-2.6$(25)[39]\,MHz & 0.01(4)[6]\{14\}\,fm$^2$ & 3.063(25)\,fm \\
\textsuperscript{24m\!}Al&   1$^+$&  4.0(28)[39]\,MHz & 0.10(4)[6]\{14\}\,fm$^2$ & 3.078(25)\,fm \\
\textsuperscript{23\!}Al & 5/2$^+$&  5.0(22)[39]\,MHz & 0.14(3)[6]\{19\}\,fm$^2$ & 3.083(33)\,fm \\
\textsuperscript{22\!}Al &   4$^+$&  2.6(51)[39]\,MHz & 0.12(7)[6]\{25\}\,fm$^2$ & 3.081(43)\,fm \\
\hline
\end{tabular}
\end{table*}

\begin{figure}[]
\includegraphics[width=0.48\textwidth]{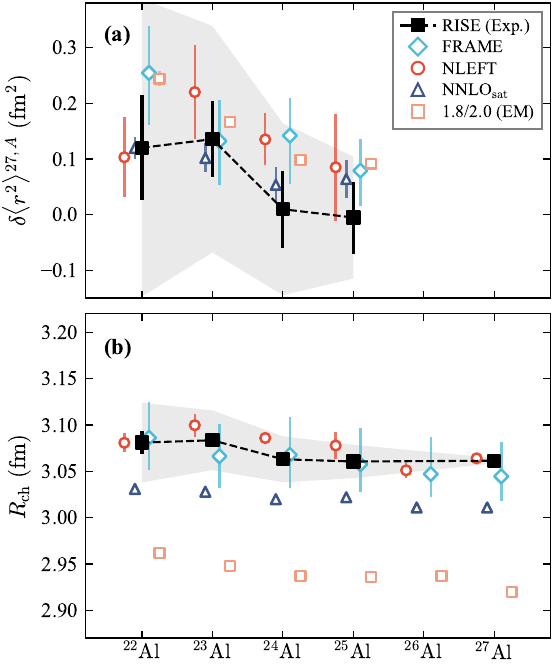}
\caption{\label{fig:diffCharge} \textbf{Differential mean-squared and absolute nuclear charge radii.} Using the atomic factors calculated in Ref.~\cite{Skripnikov2024Isotope-shift26m}, differential mean-squared charge radii relative to $^{27\!}$Al (\textbf{a}) and absolute charge radii (\textbf{b}) were extracted from the isotope shift measurements. These experimental results are compared with various ab initio methods. The experimental error bars only show the propagated statistical and systematic uncertainties from the isotope shift measurements, while the grey band represents the total uncertainty including the covariance from the atomic factors. Table~\ref{tab:Al_Shifts_Radii} shows the total uncertainty breakdown.
The error bars on VS-IMSRG predictions using the 1.8/2.0~(EM) and NNLO$_{\rm sat}$ Hamiltonians indicate model-space uncertainties,
while those on the predictions using the FRAME emulator quantify the total effective field theory and emulator uncertainties.
The error bars on Nuclear Lattice EFT (NLEFT) predictions quantify the statistical uncertainties of the Monte Carlo calculations.
}
\end{figure}

Finally, the differential mean-squared (ms) charge radii with respect to $^{27\!}$Al, $\delta\langle r^2_\mathrm{ch}\rangle^{A,27}$, could be extracted from the measured isotope shifts using the approximate relation 
\begin{equation}\label{eqn:diffChargeRadii}
    \delta\langle r^2_\mathrm{ch}\rangle^{A,27}=\frac{\delta \nu^{A,27}-K\mu^{A,27}}{F},
\end{equation}
where $F=70.11(13)$ MHz/fm$^2$ and $K=-0.7(21)$ GHz$\cdot u$ are the field- and mass-shift factors given by state-of-the-art atomic physics calculations, respectively \cite{Skripnikov2024Isotope-shift26m}, and $\mu^{A,27}\equiv 1/m^A-1/m^{27}$ is the inverse mass difference relative to the reference isotope, $^{27\!}$Al. 
The extracted nuclear charge radii results are given in Table~\ref{tab:Al_Shifts_Radii}, and compared with our theoretical results in Fig.~\ref{fig:diffCharge}. 
We compare with charge radius predictions from \textit{ab initio} nuclear structure calculations~\cite{Hergert2020,Miyagi:2025lmv} using Hamiltonians with two- and three-nucleon interactions from chiral effective field theory~\cite{Epe2009, Mac2011}.
We employ two complementary many-body methods, nuclear lattice effective field theory (NLEFT)~\cite{Elh24} and the valence-space in-medium similarity renormalization group (VS-IMSRG)~\cite{Her2016, Str2019}. 

The experimental results suggest a step-like growth in nuclear size approaching the proton drip line, with $^{23\!}$Al and $^{22\!}$Al exhibiting similar radii within experimental uncertainties. This trend does not indicate a pronounced proton-halo configuration in $^{22\!}$Al, which could be due to a dominant $d_{5/2}$ configuration for the valence protons in its ground state~\cite{Jensen2026_22Al_halo, Xu26}. {The importance of orbital angular momentum and the Coulomb interaction on the differential ms charge radius for a weakly-bound proton is discussed in the Supplementary Information.}  

NLEFT calculations employ chiral effective field theory ($\chi$EFT) interactions up to next-to-next-to-next-to-leading order (N$^3$LO) (for a review, see~\cite{Epe2009}), combined with a recently developed wavefunction-matching framework that improves convergence and mitigates Monte Carlo sign problems \cite{Elh24}. 
This approach enables precise predictions of binding energies and charge radii across light- and medium-mass nuclei using full $A$-body lattice Monte Carlo calculations. 
The low-energy constants of the three-nucleon forces are adjusted to reproduce the binding energies of select light and medium-mass nuclei. Within this framework, nuclear charge radii are obtained directly from the calculated proton-density distributions, achieving agreement with empirical data at the level of 0.03~fm. 
The same methodology was successfully benchmarked in recent measurements of silicon isotopes~\cite{Kon24}, where NLEFT reproduced absolute charge radii. 
Further progress on calculating nuclear radii within this framework was reported in~\cite{Ren25}.
Further details on the NLEFT calculations can be found in the Methods, `Nuclear Lattice Effective Field Theory'.


Our \textit{ab initio} VS-IMSRG calculations solve the $A$-body Schrödinger equation to compute the structure of nuclei starting from two- and three-nucleon interactions from chiral effective field theory~\cite{Epe2009,Mac2011}.
To probe the uncertainty of the nuclear interactions, we consider two Hamiltonians that differ in their construction and how they are fit to data: 1.8/2.0~(EM)~\cite{Heb11} and NNLO$_{\textrm{sat}}$~\cite{Eks15}.
We also employ a distribution of Hamiltonians at next-to-next-to-leading order (NNLO) in chiral effective field theory with explicit $\Delta$-isobar degrees of freedom~\cite{Jia24}.
This distribution consists of more than 8000 weighted samples constructed through history matching to determine allowed ranges for the low-energy couplings in the Hamiltonians followed by importance resampling based on the reproduction of few-body observables as well as the ground-state energy and charge radius of \textsuperscript{16}O~\cite{Jia24}. 
We propagate this distribution forward to obtain posterior predictive distributions for the charge radii of aluminium isotopes that quantify the nuclear Hamiltonian uncertainty at NNLO.
Directly solving the VS-IMSRG for all of these samples is a computationally intractable task, so
we use the newly developed nuclear emulator FRAME~\cite{Mun26}, which can predict results of full many-body calculations over complete isotopic chains. 
Trained on more than 25,000 VS-IMSRG calculations in multiple isotopic chains, the emulator reproduces the test data set consisting of VS-IMSRG predictions of absolute charge radii of aluminium isotopes with a root mean-squared error of 0.003~fm.
More details on the VS-IMSRG calculations and FRAME emulator  can be found in Methods, `VS-IMSRG and FRAME'.

In Fig.~\ref{fig:diffCharge}b, the significant difference in charge radii predicted using the 1.8/2.0~(EM) and NNLO$_{\textrm{sat}}$ Hamiltonians illustrates the strong sensitivity of charge radii to nuclear forces.
The 1.8/2.0~(EM) Hamiltonian is only optimized to few-body data and significantly underpredicts charge radii in medium-mass nuclei.
The NNLO$_{\textrm{sat}}$ Hamiltonian is additionally optimized to ground-state energies and charge radii of selected carbon and oxygen isotopes,
yielding more accurate predictions for charge radii~\cite{hag16,Soma2020}.
Both our NLEFT Hamiltonian and the distribution of NNLO Hamiltonians used by the FRAME emulator
are fit to or calibrated with ground-state properties of medium-mass nuclei.
This allows them to accurately reproduce the measured charge radii.

Finer structural details can be investigated by considering the differential mean-square charge radii,
where correlated uncertainties in neighboring isotopes largely cancel.
Interestingly, the NLEFT calculations predict a reduction in the nuclear charge radius toward the proton dripline, with $^{22\!}$Al having a smaller radius than $^{23\!}$Al. 
This behavior is different for the VS-IMSRG predictions, which consistently show a gradual increase in the radius toward the proton dripline.
 For $^{23\!}$Al, which contains two neutrons beyond $N=8$, a relative increase, or ``kink,'' in the charge radius reflects the effect of the $N=8$ shell closure, which has been observed in lighter nuclei~\cite{Gei08,Ohayon2022Na}.

\begin{figure}[t]
\centering
\includegraphics[scale=0.3]{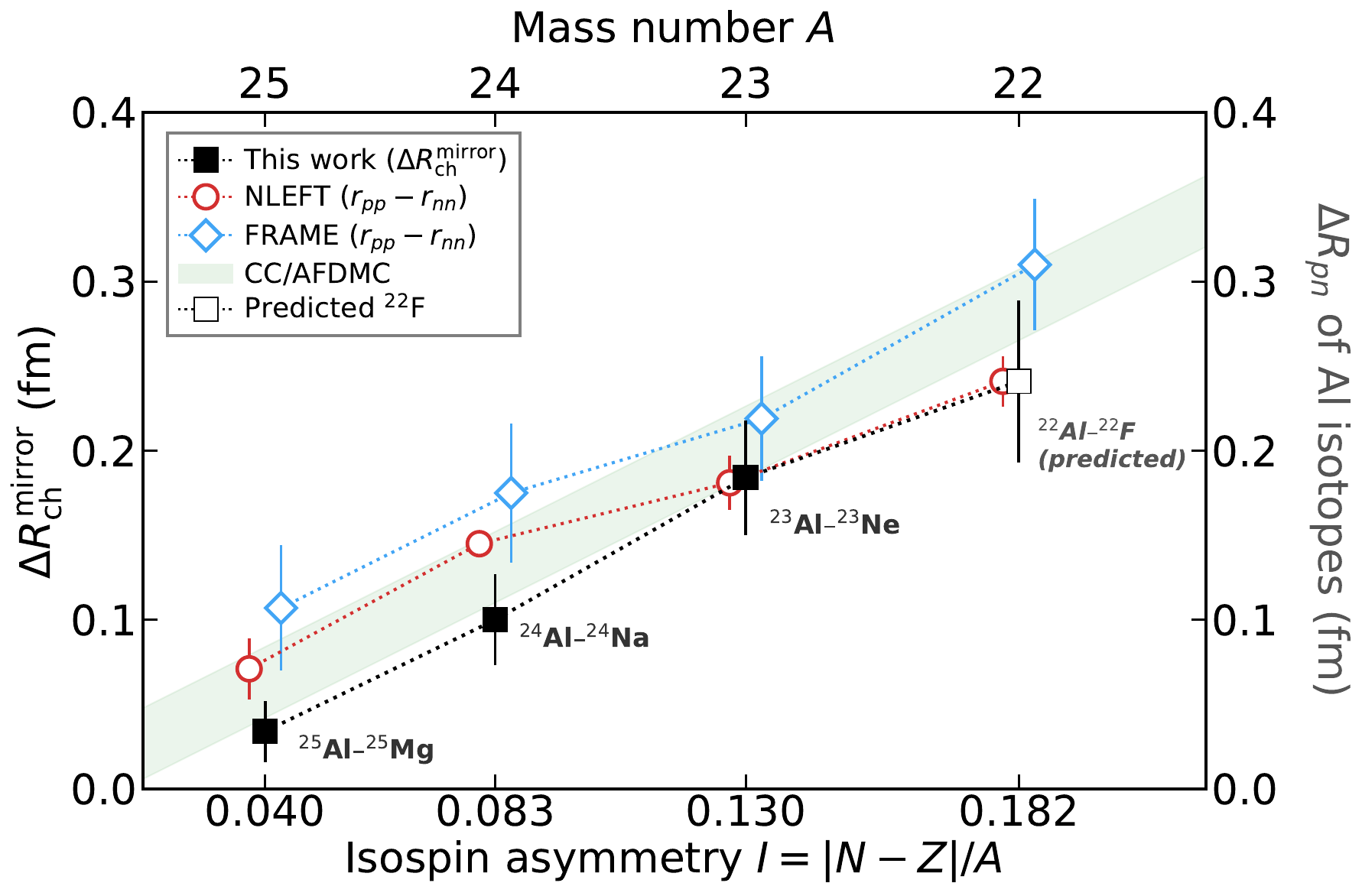}

\caption{\textbf{Mirror charge-radius differences and proton skins}. Mirror charge-radius differences 
$\Delta R_{\rm ch}^{\rm mirror}$ (left axis) compared with the 
 proton skin $\Delta R_{pn} = r_{pp} - r_{nn}$ 
(right axis) of aluminium isotopes calculated from NLEFT (red open circles) and FRAME (blue open diamonds), as a function of the isospin 
asymmetry $I$. Mirror partner radii are from 
Ref.~\cite{Ohayon2025}. The open square at $A=22$ shows the 
value of $^{22\!}$Al--$^{22}$F using our measurement $^{22\!}$Al and the predicted value of $^{22}$F from VS-IMSRG calculations, $R_{\rm ch}=2.84$~fm. The green band shows the predicted trend from CC and AFDMC 
calculations, $\Delta R_{\rm ch}^{\text{mirror}} = 1.574~I~ \pm~0.021$~fm ~\cite{Novario2023}. See text for details.}
\label{fig:proton_skin}
\end{figure}

Two mechanisms for the observed increase in charge radii towards the proton drip line are possible:
a spatially extended proton distribution could emerge, as 
expected for a proton halo;
or the proton skin could grow gradually, driven by the increasing proton-neutron asymmetry 
that characterizes well-bound nuclei.
The charge radius alone does not allow us to distinguish these.
To investigate this, a comparison between the evolution of the proton and neutron radii is needed. 
Measuring neutron radii is challenging even in stable isotopes, so the neutron radius has only been measured in \textsuperscript{27\!}Al~\cite{Androic2022PRL_Al27QWeak}.
However, under approximate charge symmetry, the difference in charge radii between mirror nuclei, $\Delta R_{\rm ch}^{\rm mirror} \equiv R_{\rm ch}({\rm Al}) - R_{\rm ch}({\rm mirror})$, provides a direct experimental proxy for the proton skin, $\Delta R_{pn} \equiv r_{pp} - r_{nn}$, i.e.~the difference between 
the point-proton and point-neutron root-mean-square 
radii~\cite{Brown2017,Yang2018}. Using our new experimental results together with the mirror partner radii $R_{\mathrm{ch}}(^{25}\mathrm{Mg}) = 3.026(3)\,\mathrm{fm}$, $R_{\mathrm{ch}}(^{24}\mathrm{Na}) = 2.963(11)\,\mathrm{fm}$, and $R_{\mathrm{ch}}(^{23}\mathrm{Ne}) = 2.899(10)\,\mathrm{fm}$ from Ref.~\cite{Ohayon2025}, we can extract the charge-radius differences between mirror nuclei shown with black filled 
squares in Fig.~\ref{fig:proton_skin} (left axis) as a function of the isospin asymmetry $I = |N - Z|/A$. On the right axis, the same scale is 
used to display the proton skin $\Delta R_{pn}$ of the aluminium 
isotopes, as computed by the two \textit{ab initio} approaches 
discussed in this work: NLEFT (red open circles, from Table~\ref{tab:NLEFT_radii}) 
and FRAME (light-blue open diamonds, with the neutron radius calculated from the NNLO$_{sat}$ Hamiltonian). The green band shows the slope obtained from coupled cluster (CC) and auxiliary field diffusion Monte Carlo (AFDMC) calculations of light and medium-mass nuclei~\cite{Novario2023}.

Both the experimental mirror charge-radius differences and the predicted proton skins of aluminium isotopes follow the trend expected for well-bound nuclei as a function of isospin asymmetry.
This disfavors a proton-halo structure for these isotopes. If a 
proton halo were present in $^{22\!}$Al or $^{23\!}$Al, the valence 
proton would extend well beyond the neutron core, producing a 
proton skin significantly larger than the systematic trend 
established by well-bound mirror pairs~\cite{Novario2023,Ohayon2025}. 
The mirror charge-radius difference would then appear as a clear outlier above the linear trend in Fig.~\ref{fig:proton_skin}.
An extension of Fig.~\ref{fig:proton_skin} to other nuclei, including the inclusion of a two-proton halo nucleus, is shown in Methods, `Mirror nuclei and proton skin'.

\section{Conclusions and Perspectives}
We present the first laser-spectroscopy measurements of the nuclear charge radii of aluminium isotopes from stable $^{27\!}$Al to the proton-drip-line nucleus $^{22\!}$Al. These results were enabled by the newly commissioned RISE instrument at the BECOLA facility at FRIB, which demonstrated high-sensitivity measurements of isotopes produced at the limits of nuclear existence. The combined capabilities of RISE and FRIB will enable precision laser spectroscopy measurements of short-lived nuclei that were previously inaccessible.

Our results show a correlation between the proton skin and mirror charge-radius differences, with the observed charge radii of Al isotopes following the pattern established by well-bound nuclei. A proton halo in $^{22,23\!}$Al, as suggested by earlier studies~\cite{Cai2002,Lee2020,Campbell2024}, would require the proton distribution to extend significantly beyond the neutron core, yielding a mirror shift well in excess of the systematic trend. Our results rule out such an enhancement at the present level of precision. These findings are supported by two complementary \textit{ab initio} nuclear calculations: NLEFT calculations reproduce the absolute scale of the measured charge radii, with a predicted reduction from $^{23\!}$Al to $^{22\!}$Al, consistent with the effect expected for the shell closure at $N=8$. In contrast, VS-IMSRG calculations exhibit a monotonic increase of the radii toward the drip line.

Our developments open a path towards the systematic exploration of charge radii in the most neutron deficient isotopes of elements up to and beyond $Z = 14$. Extending laser spectroscopy measurements to neighboring elements, including Si~($Z=14$), P~($Z=15$), and S~($Z=16$), which can now be produced at FRIB, will allow a comprehensive understanding of shell effects, deformation, and halo formation in medium-mass nuclei. The combined capabilities of RISE at BECOLA and FRIB now make precision laser spectroscopy measurements possible in regions of the nuclear chart where electromagnetic, strong, and continuum effects intertwine most strongly. The aluminium chain represents the first milestone in this program, with exotic isotopes with extreme proton-to-neutron ratios across the nuclear chart now within reach, motivating future studies of nuclear size at the limits of stability.\\

{\footnotesize
\noindent\textbf{Acknowledgments}
This material is based upon work supported by the U.S. Department of Energy, Office of Science, Office of Nuclear Physics and used resources of the Facility for Rare Isotope Beams (FRIB) Operations, which is a DOE Office of Science User Facility under Award Number DE-SC0023633. The RISE experiment was supported by the Office of Nuclear Physics, U.S. Department of Energy, under grants DE-SC0021176, DE-SC0021179 and DE-SC0000661. We acknowledge support from the National Science Foundation, grant No. PHY-21-11185, and by the Deutsche Forschungsgemeinschaft (DFG, German Research Foundation) -- Project-ID 279384907 -- SFB 1245. A.J.B. acknowledges support from the NSF AGEP Fellowship 6949423. We are grateful for the technical support provided by the Bates Laboratory at MIT, and in particular we thank Ernest E. Ihloff for his invaluable assistance during the construction and installation of the RISE instrument. We also extend gratitude to John McGlashing and Joseph D Cucinotta for their tireless assistance at MIT, as well as for transporting much of the RISE equipment to FRIB. J.K. acknowledges support from a Feodor-Lynen postdoctoral research fellowship funded by the Alexander-von-Humboldt Foundation. S.E.C. acknowledges support from the DOE NNSA SSGF under DE-NA0003960. C.M.I. acknowledges support from the MSU ASET Traineeship under the DOE award no. DE-SC0018362. 
P.A.~acknowledges support by the European Union under the Marie Skłodowska-Curie grant agreement No.~101152722. Views and opinions expressed are however those of the author(s) only and do not necessarily reflect those of the European Union or the European Research Executive Agency (REA). Neither the European Union nor the granting authority can be held responsible for them.
 A.B. acknowledges the support of the Natural Sciences and Engineering Research Council of Canada (NSERC) [PDF-587464-2024].
We are grateful for discussions with the NLEFT Collaboration, and support acknowledgements for NLEFT Collaboration members are as follows: S.E. [Scientific and Technological Research Council
of Turkey (TUBITAK project no. 123F464)]; D.L. and Y.-Z.M [U.S. Department of Energy grants DE-S0013365, DE-SC0023175, DE-SC0026198, Oak Ridge Leadership Computing Facility computing resources through the INCITE award “Ab-initio
nuclear structure and nuclear reactions”, National Energy Research Scientific Computing Center (NERSC) using NERSC award NP-ERCAP0036535, the Advanced Cyberinfrastructure Coordination Ecosystem: Services \& Support (ACCESS) program through allocation PHY250148]; U.-G.M. and S.Z. [EXOTIC grant from the European Research Council (ERC) under the European Union’s Horizon
2020 research and innovation programme (grant agreement
No. 101018170), U.-G.-M. [CAS President’s International Fellowship Initiative (PIFI) (Grant No. 2025PD0022), Gauss Centre for Supercomputing
e.V. (www.gauss-centre.eu) for computing time on the GCS Supercomputer JUWELS
at J\"ulich Supercomputing Centre (JSC) and on HoreKa by the National
High-Performance Computing Center at KIT (NHR@KIT)]; T.W. [NSFC of
China under Grants No. 12125501 and No. 12550007]; R.B.Y. [National Science Foundation Award No. 2137718].
The work of M.H. was supported 
by the Laboratory Directed Research and Development Program of Oak Ridge National Laboratory, managed by UT-Battelle, LLC, for the U.S.\ Department of Energy
and by the U.S.\ Department of Energy, Office of Science, Office of Advanced Scientific Computing Research and Office of Nuclear Physics, Scientific Discovery through Advanced Computing (SciDAC) program (SciDAC-5 NUCLEI).
This research used resources of the Oak Ridge Leadership Computing Facility located at Oak Ridge National Laboratory, which is supported by the Office of Science of the Department of Energy under contract No.~DE-AC05-00OR22725.
M.H. gratefully acknowledges the Gauss Centre for Supercomputing e.V.\ (www.gauss-centre.eu) for funding this project by providing computing time through the John von Neumann Institute for Computing (NIC) on the GCS Supercomputer JUWELS at Jülich Supercomputing Centre (JSC).

\noindent\textbf{Author Contributions}
A.R.V., R.F.G.R and K.M. are the spokespeople of the experiment. B. J. R., A. D., K. M., M. D. M., A. O.C., and H. S. prepared and maintained the laser spectroscopy apparatus locally with contributions from MIT personnel; A.R.V., S.G.W., and A.J.B..
X.C., N.D.G., C.I., and C.S. developed and provided stopped beams of radioactive Al isotopes for the experiment. A.J.B., B.J.R., S.E.C., A.D., H.E., R.F.G.R., C.M.I., C.M.J., F.M.M., K.M., M.D.M., A.O.C., J.P., S.P., F.C.P.C., R.R., H.S., S.G.W., and R.B.Y. conducted the measurements during the beam time. A.J.B. and B.J.R. performed the data analysis. P.A., A.B., M.H., and J.M.M. performed IMSRG theoretical calculations. S.E., D.J.L., Y.Z.M., U.G.M., T.W., and S.Z. performed NLEFT theoretical calculations. A. J. B. and R. F. G. R. wrote the initial manuscript with input from B. J. R., A. B., D. J. L., K. M., and C. S.. A.J.B, J.M.M., R.F.G.R.,and A.B prepared the figures with input from all co-authors. R.F.G.R. and K.M. supervised the project and secured funding. All authors discussed the results, contributed to the interpretation of the data, and reviewed and edited the manuscript. A.J.B. and B.J.R. contributed equally to this work.

\noindent\textbf{Competing Interests}
The authors declare no competing interests.

\noindent\textbf{Data Availability}
The data that support the findings of this study are available from the
corresponding author upon request.
}
\bibliographystyle{naturemag}
\bibliography{references}

\section*{Methods}

\subsection*{Laser system}
\begin{figure}[h!]
\includegraphics[width=0.5\textwidth]{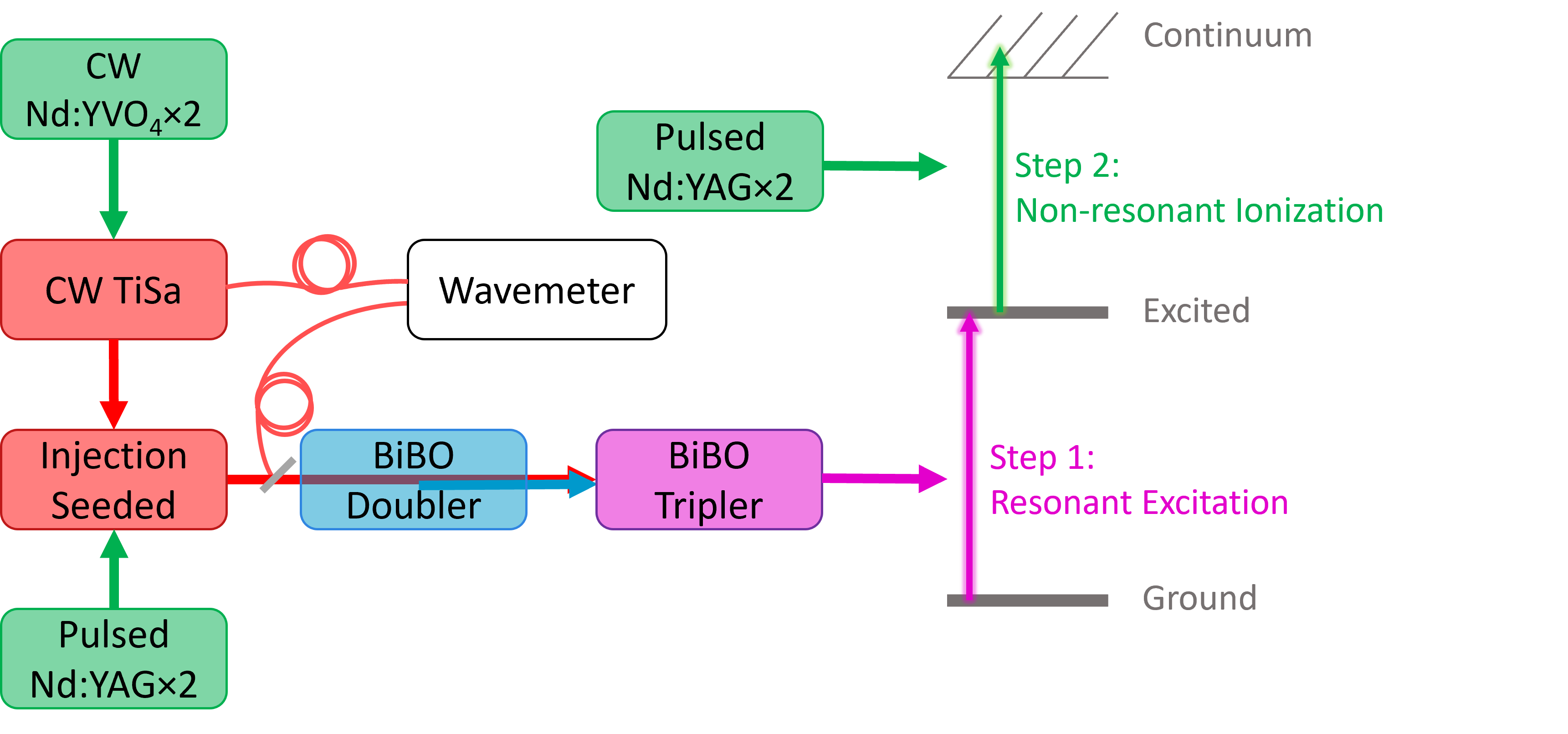} 
\caption{\label{fig:lasers} \textbf{Laser system.}
A continuous wave (CW) Titanium:Sapphire (Ti:Sa) laser system (Matisse), pumped by a CW frequency-doubled neodymium-doped yttrium orthovanadate (Nd:YVO$_4$) laser (Millennia) was used to seed an injection-seeded Ti:Sa cavity pumped at 10\,kHz, producing high-energy laser pulses locked to the Matisse frequency. Single-pass BBO doubling and BiBO tripling yielded $\sim265~$nm light for resonant excitation, while a frequency-doubled 100\,Hz neodymium-doped yttrium aluminium garnet (Nd:YAG) laser (Merion) supplied the ionization pulse.
}
\end{figure}

A schematic diagram of the laser systems and the scheme employed during the experiment are shown in Fig.~\ref{fig:lasers}. A continuous-wave (CW) frequency-doubled neodymium-doped yttrium orthovanadate  laser (Spectra Physics Millennia) pumped a CW single-mode ring Titanium:Sapphire (Ti:Sa) laser (Spectra Physics Matisse TS), which was used to seed and frequency-lock an injection-seeded Ti:Sa cavity. The injection-seeded Ti:Sa was pumped by a 10\,kHz frequency-doubled neodymium-doped yttrium aluminium garnet laser (Photonics TU-H). This configuration produces high-energy Ti:Sa pulses with frequencies set by the Matisse output. The pulses are frequency doubled and tripled in single-pass BBO and BiBO crystals to generate $\sim265$\,nm light for resonant excitation of the $3s^23p\,\,^2P_{1/2} \rightarrow 3s^25s\,\,^2S_{1/2}$ transition. The fundamental frequencies of both Ti:Sa lasers were monitored by a wavemeter (High Finesse WSU-30). A frequency-doubled 100\,Hz Nd:YAG laser (Quantel Merion MW 7-100) provides the selective ionization pulses.

\subsection*{Measured spectra and data analysis}
\label{sec:data}
\begin{figure}[h]
\includegraphics[width=0.5\textwidth]{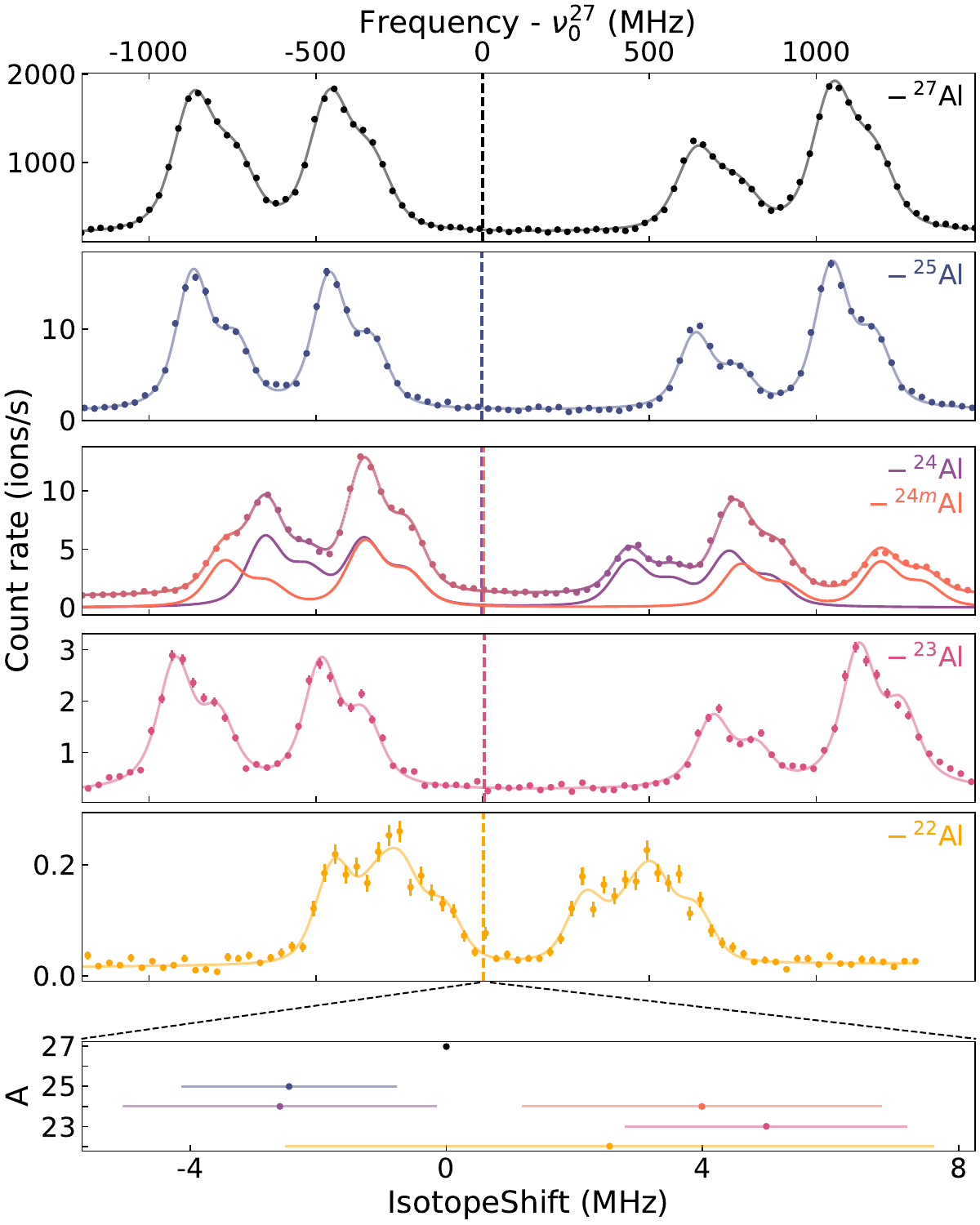} 
\caption{\label{fig:iso2d} \textbf{Measured hyperfine structure spectra.}
Top: Spectra of the $3s^23p\,\,^2P_{1/2} \rightarrow 3s^25s\,\,^2S_{1/2}$ transition in $^{27, 25, 24, 23, 22\!}$Al. For each spectrum, the structure is split into 4 primary peaks due to the hyperfine interaction between the nuclear spin and the total electronic angular momentum. For each of these primary peaks, there is a corresponding satellite peak due to energy transfer in the charge-exchange process \cite{Dockery2025SpectralExchange}, which appears at higher energy for the present case of anticollinear laser and Al-beam propagation. The spectrum for $^{24\!}$Al also contains an isomeric state produced during the fragmentation process.
Bottom: Isotope shift values and uncertainties for each isotope, obtained by extracting centroids from each of the above spectra and subtracting the centroid for $^{27\!}$Al. For $^{24\!}$Al, the spectrum was Doppler shifted using the ground and isomer masses to extract the respective centroids.}
\end{figure}

Fig.~\ref{fig:iso2d} shows the measured hyperfine spectra for $^{27,25-22\!}$Al on the $3s^23p\,\,^2P_{1/2} \rightarrow 3s^25s\,\,^2S_{1/2}$ transition. The extracted centroids were then used to determine the isotope shifts relative to $^{27\!}$Al, as shown in the lower panel.

For the $3s^23p\,\,^2P_{1/2} \rightarrow 3s^25s\,\,^2S_{1/2}$ transition in aluminium, there is a characteristic spectrum of 4 primary peaks, due to the hyperfine interaction. For a transition centroid $\nu_0$, and hyperfine constants for the ground and excited states, $\mathcal{A}_{3P}$ and $\mathcal{A}_{5S}$, respectively, the transition from a ground total angular momentum $F=I+J$ to excited $F'=I+J'$ occurs at a frequency given by:
\begin{equation}
    \nu_{FF'} = \nu_0 + \frac{1}{2}\left(\mathcal{A}_{5S}K_{F'IJ'}-\mathcal{A}_{3P}K_{FIJ}\right),
    \label{eqn:transitionFreqs}
\end{equation}
where $K_{FIJ}\equiv F(F+1)-I(I+1)-J(J+1)$.
The lineshapes of these individual resonances are determined by a combination of natural linewidth, Doppler broadening, and beam energy jitter, and are well described by a Voigt profile:
\begin{equation}
    L(\delta_{F,F'};\gamma, \sigma) = \int_{-\infty}^{\infty}\frac{\Gamma}{\pi(x^2+\Gamma^2)}\frac{1}{\sqrt{2\pi}\sigma}e^{-(\delta_{F,F'}-x)^2/2\sigma^2} dx,
    \label{eqn:lineshape}
\end{equation}
where $\Gamma$ and $\sigma$ represent the broadening terms of Lorentzian and Gaussian nature, respectively, and $\delta_{F,F'} \equiv \nu-\nu_{FF'}$ is the detuning from resonance.
For each of these primary peaks, there also exist corresponding satellite peaks (sp) due to energy-exchange processes in the CEC during neutralization \cite{Bendali1986Na+-NaSpectroscopy, Klose2012TestsSpectroscopy}. The character of the satellite peak is mainly determined by the ion bunch beam energy, the species of alkali used for neutralization, and the electronic energy structure of the beam element~\cite{Dockery2025SpectralExchange}. These satellite peaks were accommodated by introducing additional peaks to the model, scaled and shifted by free parameters $p_{\mathrm{sp}},\delta_{\mathrm{sp}}$, respectively.

Summing the contributions from all primary and satellite peaks, as well as a linear background term, the total spectrum for a single nuclear species, is given by
\begin{equation}
\label{eqn:lineshape}
S(\nu) = b + k\nu
+ \sum_{FF'} h_{FF'} \times \left(
\begin{aligned}
    &\, L\!\left(\delta_{FF'};\gamma, \sigma\right) \\
    +  p_{\mathrm{sp}} &\,
      L\!\left(\delta_{FF'}-\delta_{\mathrm{sp}};\gamma, \sigma\right)
\end{aligned}
\right),
\end{equation}
where $h_{FF'}$ is the overall amplitude for the hyperfine transition from $F$ to $F'$.
This amounts to 13 free parameters: 2 for the linear background ($b$ and $k$), 4 for the peak heights ({$h_{FF'}$}), 2 for the Voigt lineshape ($\gamma$ and $\sigma$),
2 for describing the satellite peaks ($p,\nu_{\mathrm{sp}}$),
and 3 to describe the resonant frequencies ($\nu_0$, $\mathcal{A}_{3P}$, and $\mathcal{A}_{5S}$).
Since the same electronic transition was measured for each isotope, the ratio of hyperfine constants, $\alpha_{\mathcal{A}}^{(A)}\equiv \mathcal{A}^{(A)}_{3P}/\mathcal{A}^{(A)}_{5S}$, should be the same for every isotope, up to the level of a differential hyperfine anomaly. This ratio was measured for $^{27\!}$Al, and then used to constrain the fits for the rare isotope spectra.
For the case of $^{24\!}$Al, 6 additional parameters were needed to capture the spectrum, corresponding to the isomer peak heights, one hyperfine constant, and the isomer's centroid.

\begin{figure}[h]
\includegraphics[width=0.5\textwidth]{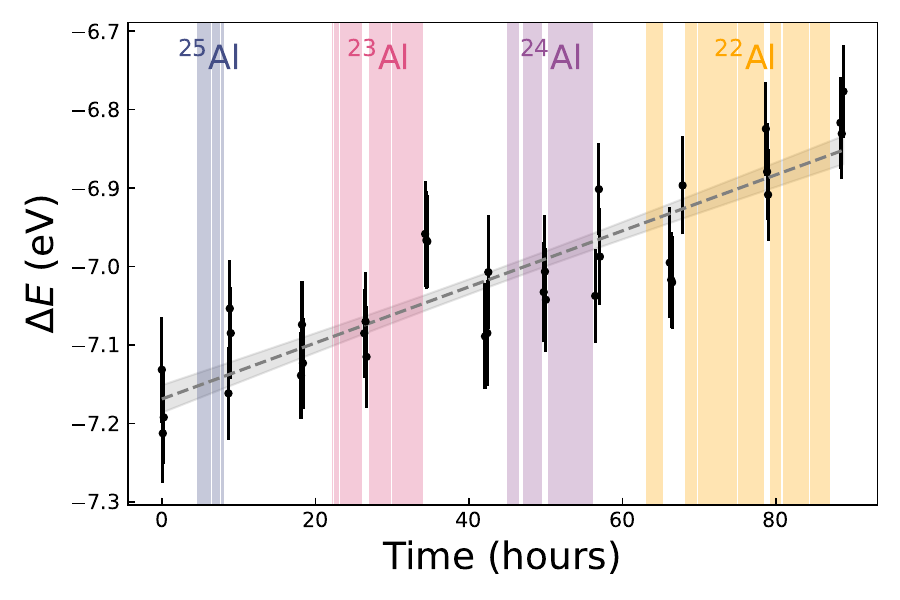} 
\caption{\label{fig:calibration} \textbf{Beam-energy calibration.}
The black points show the measured beam energy deviations over time, with associated error bars, calculated from centroid measurements on $^{27\!}$Al spectra taken regularly throughout the experiment. The dashed line indicates a linear best fit for the beam energy correction vs time, which was applied to the unstable isotope data. The colored bands display the measurement intervals for each unstable isotope.
}
\end{figure}

The production yields and measurement times for the aluminium isotopes studied in this work are summarized in Table~\ref{tab:Production}. The beam intensities were determined using a Si surface-barrier detector located before the BECOLA RFQCB. The ions were stopped in an aluminium foil placed in front of the detector, and the emitted $\beta$ particles were detected, corresponding to a solid-angle coverage of approximately 30\%. For $A=24$, both the ground and isomeric states were populated. In the absence of independent yield information, equal production of the two states was assumed
. Throughout the experiment, $^{27\!}$Al hyperfine spectra were collected as reference spectra for a beam energy calibration. Fig.~\ref{fig:calibration} shows the measurement times and beam-energy deviations extracted from the reference-isotope measurements, as well as the measurement intervals for all short-lived isotopes. Using a linear fit of the calibration data in Fig. \ref{fig:calibration} and the timestamps of the individual voltage steps across each spectrum, a correction was applied to the beam energies for all short-lived isotopes. Further details on this beam energy correction procedure can be found in \cite{Konig2021BeamSpectroscopy, Brinson2026RISE}.

\begin{table}[h]
    \caption{\textbf{Isotope masses, production yields, and measurement times.} The rates of Al isotopes were measured using a Si surface barrier detector located right before the injection into the BECOLA RFQ cooler/buncher. The beams were stopped in an aluminium foil placed in front of the detector, and the emitted $\beta$ decays were detected, covering a solid angle of approximately 30\%. Values with an asterisk indicate beams which had magnesium contamination. The quoted rates are estimates of the amount of aluminium; see Ref. \cite{rickey2025halo} for details. The mass values of $^{22\!}$Al and $^{23-27\!}$Al were taken from Ref.~\cite{Campbell2024} and \cite{AME2020}, respectively. The mass for $^{24\mathrm{m}\!}$Al was obtained using the 425.8(1)\,keV excitation energy of the isomeric state~\cite{Honkanen1979Decay24Mg}. For $A=24$, the total measured yield was 28\,000\,cps between both the ground and isomeric states.}
    \centering
    \begin{tabular}{l||l|r|r}
        A & Mass (u) & Yield (pps) & Meas. time (h)\\
        \hline
        \rule{0mm}{3mm}25 &24.99042831(7)   & 36000 & 3.3  \\
        24 &23.99994760(24)  & 28000 & 9.9\\
        23 &23.0072444(4)    & 19199* & 10.6\\
        22 &22.01942311(30) & 1102* & 19.7 \\
    \end{tabular}
    \label{tab:Production}
\end{table}

By fitting the energy-corrected spectra for each isotope, the isotope shifts were derived from the best-fit estimates on the centroids as $\delta\nu^{A,27} = \nu_0^{A}-\nu_0^{27}$. These shifts are given in Table \ref{tab:Al_Shifts_Radii}.

\subsubsection{Systematic uncertainties}
The dominant sources of systematic uncertainty are enumerated below:
\paragraph{Wavemeter uncertainty} Unlike the case of the absolute centroid discussed in \cite{Brinson2026RISE}, most of the WSU-30 wavemeter uncertainty cancels out for isotope shift measurements, since most of this systematic uncertainty is found to be highly correlated, and reproducible over time \cite{Konig2020OnDeterminations}. The uncertainty that does not cancel for difference measurements is due to the local error structure across the free spectral range of the wavemeter's highest resolution interferometer (3~MHz), and temporal drift not compensated for by autocalibration (0.05~MHz). Since the centroid for $^{27\!}$Al was determined from anticollinear and collinear spectroscopy measurements at two distinct frequency setpoints, the non-canceling uncertainty contribution for this isotope is 2.1~MHz, coming from uncertainty propagation of the formula $\nu_0=\sqrt{\nu_a\nu_c}$, where $\nu_{a/c}$ represent the centroids measured with anti/collinear spectroscopy, respectively. All other isotopes were only measured at a single frequency setpoint, and thus contribute 3~MHz each. The total wavemeter uncertainty for each isotope shift measurement is therefore $\sigma_{\delta\nu^{A,27}}^{(\mathrm{wm})} = \sqrt{\sigma^2_{\nu^{27}}+\sigma^2_{\nu^{A}}}=3.7$~MHz.

\paragraph{Random variation} It is possible for the recorded centroids to be affected by time-varying systematics. Examples of such systematics include frequency drift of the helium-neon laser used for calibrating the wavemeter, ambient fluctuations of the voltage divider, and imperfect realignment of the resonant excitation laser with the path of the aluminium atom bunches.
Since the reference isotope was measured throughout the experiment, the centroid scatter from the energy corrected calibration measurements can be used to estimate the total uncertainty from all time-varying sources.
By grouping the calibration results from each realignment and comparing the observed variance of the calibration centroids (1.2~MHz$^2$) with the expected variance from the centroids' statistical uncertainties (0.5~MHz$^2$), a residual systematic drift of 0.85~MHz was deduced. 
For the isotope shifts, this uncertainty should be added in quadrature, contributing $\sqrt{2}*0.85\approx 1.25$~MHz of systematic uncertainty for all isotopes considered.

\paragraph{Beam-energy correction}By evaluating the energy calibration function uncertainty at the average scan time for each isotope, a characteristic beam energy uncertainty is determined. This beam energy uncertainty ranges from 10-16~mV for the rare isotope scans, translating to 0.3-0.48 MHz uncertainty in the rare isotope centroids. A 3.6~mV uncertainty in the anti/colinear analysis for $^{27\!}$Al \cite{Brinson2026RISE} also contributes a 0.1 MHz uncertainty on the reference centroid.

\paragraph{A-ratio uncertainty} The measured ratio between the ground and excited state hyperfine structure constants for $^{27\!}$Al was measured to be $\alpha_A^{27}\equiv\frac{\mathcal{A}^{27}_{3P}}{\mathcal{A}^{27}_{5S}}=$ 3.684(2), and was used to constrain the fits for the unstable isotopes.
To account for the uncertainty in $\alpha_A$, centroids were also extracted from fits sampling 1$\sigma$ around the best value. The resulting centroid shifts were found to be below 0.06~MHz for all isotopes. We note that we similarly checked the dependence of the isotope and isomer shifts on a possible HFS anomaly, assuming an anomaly as large as 1\%. Even in this extreme case, the change of the isomer shift was only about half of its statistical uncertainty and the effect was smaller in all other reported shifts. 

\subsection*{Nuclear Lattice Effective Field Theory (NLEFT)}

Our \textit{ab initio} lattice simulations were performed using a high-fidelity chiral N$^3$LO wavefunction matching Hamiltonian \cite{Elh24}.
We employed a lattice spacing of $a=1.32$ fm and a box size of $L^3=10^3$. 
All lattice results were obtained after Euclidean-time extrapolation, ensuring convergence to the true ground state.
Following the strategy of our previous work \cite{Kon24}, we fit the three-nucleon coefficients $c_E^{(l)}$ and $c_E^{(t)}$ to ensure good agreement with the binding energies of the aluminium isotopic chain.
Proton and neutron radii were computed on the lattice using the pinhole method \cite{Elh17, Shu25, Ren25}. 
Higher-order corrections from chiral interactions were included perturbatively. 
Charge radii were obtained from point-proton radii $r_{\mathrm{pp}} $ using the standard formula: $\left\langle r_{\mathrm{ch}}^2\right\rangle=\left\langle r_{\mathrm{pp}}^2\right\rangle+R_{\mathrm{p}}^2+\frac{N}{Z} R_{\mathrm{n}}^2+\frac{3 \hbar^2}{4 m_{\mathrm{p}}^2 c^2}$, where $\frac{3 \hbar^2}{4 m_{\mathrm{p}}^2 c^2} \approx 0.033\, \mathrm{fm}^2$, $R_{\mathrm{n}}^2=-0.105_{-0.006}^{+0.005}\, \mathrm{fm}^2$ \cite{Fil21} and $R_{\rm p}=0.840_{-0.002-0.002}^{+0.003+0.002} \,\mathrm{fm}$ \cite{Lin22}.

\begin{table}[h]
\centering
\caption{NLEFT calculation for charge, proton, and neutron radii, units in fm. \label{tab:NLEFT_radii}}
\begin{tabular}{clll}
\hline
Isotope &
$\langle r_{\rm ch} \rangle$ &
$\langle r_{\rm pp} \rangle$ &
$\langle r_{\rm nn} \rangle$ \\
\hline
\rule{0mm}{3.5mm}$^{22\!}$Al  & 3.081(10) & 2.971(10) & 2.730(11) \\ 
$^{23\!}$Al  & 3.100(12) & 2.991(12) & 2.810(11) \\ 
$^{24\!}$Al  & 3.086(5)  & 2.979(5)  & 2.834(5)  \\ 
$^{25\!}$Al  & 3.078(15) & 2.972(15) & 2.901(10) \\ 
\hline
\end{tabular}
\end{table}

A notable feature of the aluminium charge radii is a pronounced kink between $^{22-24\!}$Al.
In general, charge radii are expected to increase toward the proton drip line due to the reduced strength of attractive $pn$ correlations.
While this overall trend is observed, $^{23\!}$Al stands out among the monotonic behavior, exhibiting an enhanced charge radius, as also seen in Fig~\ref{fig:figure_rch}. 
A similar pattern was reproduced in our lattice calculations. 
Some relevant data are summarized in Tab~\ref{tab:NLEFT_radii}.
A possible origin of the $^{23\!}$Al charge radius kink is neutron pairing.

As the mass number decreases from $A=25$ to $A=22$, the neutron radius generally decreases; however, $^{23\!}$Al shows a distinctly larger value than average: $r_{\mathrm{nn}}(A=23) - [r_{\mathrm{nn}}(A=22)+r_{\mathrm{nn}}(A=24)]/2=0.028$.
This suggests that the two neutrons above the $N=8$ closed shell form a neutron pair and can spatially move more freely than the remaining nucleons.
Thus, the resulting shift of the nuclear center of mass leads to an enhancement of both the proton and charge radii.

\subsection*{VS-IMSRG and FRAME}


The valence-space in-medium similarity renormalization group~\cite{Tsu2012,Bog2014,Her2016,Str2016,Str17,Str2019} is an \textit{ab initio} many-body method that solves the many-body Schrödinger equation for nuclear Hamiltonians
with two- and three-nucleon potentials from chiral effective field theory~\cite{Epe2009,Mac2011}.
From a Hartree-Fock state, it computes a unitary transformation that decouples a core and valence space from the remaining Hilbert space.
This transformation is truncated at the normal-ordered two-body level, the VS-IMSRG(2) approximation.
The resulting effective valence-space Hamiltonian is diagonalized using \textsc{kshell}~\cite{Shi19}, giving the ground-state energy.
Other ground-state observables are evaluated by consistently transforming the associated operator and evaluating its expectation value in the decoupled valence space.

We compute charge radii starting from the point-proton radius operator, including finite-nucleon-size, spin-orbit, and Darwin-Foldy corrections~\cite{Friar1997PRA_DarwinFoldy, Ong2010PRC_SpinOrbit, Heinz:2024juw}: $\left\langle r_{\mathrm{ch}}^2\right\rangle=\left\langle r_{\mathrm{pp}}^2\right\rangle + \left\langle r_{\mathrm{so}}^2\right\rangle +R_{\mathrm{p}}^2+\frac{N}{Z} R_{\mathrm{n}}^2+\frac{3 \hbar^2}{4 m_{\mathrm{p}}^2 c^2}$ with $R_{\mathrm{n}}^2=-0.116\,\mathrm{fm}^2$  and $R_{\rm p}^2=0.707\,\mathrm{fm}^2$~\cite{Workman.2022}.
The value used for $R_{\mathrm{n}}^2$ differs from that used in NLEFT calculations, but this difference is negligible for this study.
Spin-orbit corrections, which are not included in NLEFT calculations, range from $\approx 0.00\:\mathrm{fm}^2$ in $^{27\!}$Al to $\approx 0.05\:\mathrm{fm}^2$ in $^{22\!}$Al.
These corrections are small compared to the Monte Carlo sampling uncertainty in our NLEFT calculations and compared to the Hamiltonian uncertainty quantified using the FRAME emulator.

Our calculations are expanded in a spherical harmonic oscillator basis with oscillator frequency $\hbar \omega = 16\,\mathrm{MeV}$,
truncated to include states with $2n + l \leq e_{\rm max} = 12$ (with the principal quantum number $n$ and the orbital angular momentum $l$). 
Three-body potential matrix elements are truncated at $E_\mathrm{3max}=24$~\cite{Miy2022}, sufficient for convergence up to tin isotopes.
For our calculations of aluminium isotopes, we employ an $^{16}$O core and an $sd$-shell valence space.

Nuclear Hamiltonians are uncertain.
We use several Hamiltonians with two- and three-nucleon interactions from chiral effective field theory to probe this uncertainty.
They differ in their construction and how they are fit to data.
The 1.8/2.0~(EM) Hamiltonian~\cite{Heb11} is optimized to nucleon-nucleon scattering data and ground-state properties of nuclei with $A\leq 4$.
It accurately predicts ground-state energies and spectra across the nuclear chart but severely underpredicts charge radii~\cite{Simonis2017}.
The 1.8/2.0~(EM7.5) and 1.8/2.0~(sim7.5) Hamiltonians~\cite{Art24} are constructed in a similar manner but additionally the short-range low-energy constants of the three-nucleon potential were optimized to the ground-state energy and charge radius of \textsuperscript{16}O.
The NNLO$_{\rm sat}$ Hamiltonian~\cite{Eks15} is similarly optimized to the ground-state energies and charge radii of $^{14}$C and $^{16}$O and additionally the ground-state energies of $^{22,24,25}$O.
The $\Delta {\rm NNLO}_{\rm GO}$ Hamiltonian~\cite{Jia20} is optimized to nuclear matter properties instead of medium-light nuclei.
All of the Hamiltonians optimized to nuclei $A> 4$ or nuclear matter yield significantly more accurate predictions of charge radii of medium-mass nuclei~\cite{Tews2020}.

We also employ a distribution of Hamiltonians at next-to-next-to-leading order in chiral effective field theory with explicit $\Delta$-isobar degrees of freedom~\cite{Jia24}.
This distribution comprises over 8000 weighted samples with varying low-energy constants.
The allowed combinations of low-energy constants are determined through an iterative history matching procedure with based nucleon-nucleon scattering data and the properties of $A=2$, 3, and 4 nuclei.
The relative weight of each interaction is determined from a likelihood based on the reproduction of few-body observables as well as the ground-state energy and charge radius of \textsuperscript{16}O.

Directly solving the VS-IMSRG for this distribution of Hamiltonians
for all aluminium isotopes of interest would be a computationally intractable task.
Instead we train and employ the newly developed FRAME emulator~\cite{Mun26}, which allows us to accurately sample nuclear charge radii predictions for thousands of values of the LECs at a fraction of the time and cost of a full many-body calculation. FRAME combines parametric matrix models~\cite{Coo24} with the multi-fidelity architecture and multi-isotope encoder of the BANNANE emulator~\cite{Bel26}. As such, it preserves the Hamiltonian structure, enabling accurate predictions even when extrapolating to unseen parameters, emulates over different nuclei simultaneously, and can be trained using mostly inexpensive calculations in small model spaces (with $e_{\rm max}$ significantly smaller than 12). 


The FRAME emulator is trained using the set of 8188 nuclear Hamiltonians discussed above. 
More than 25000 training samples were generated along the oxygen, aluminium, and silicon isotopic chains using VS-IMSRG calculations at different model-space truncations $e_{\rm max}$ (which serve as different fidelities in the model).
Table~\ref{tab:FrameTraining} shows the number of data points used for each isotope and fidelity. Most notably, the vast majority of the training data is taken at low fidelity, meaning that inexpensive calculations are sufficient to train the model.

\begin{table}[]
\caption{\textbf{Number of training samples per isotopes and fidelity for the FRAME emulator.} The number of samples used to train the FRAME emulator for each isotope and each fidelity level (given by the model-space truncation $e_{\rm max}$). The samples are taken randomly from the set of $\sim$8000 Hamiltonians from Ref.~\cite{Jia24}. We see that only few samples of the most expensive calculations (i.e., $e_{\rm max} = 10$) are needed to obtain reliable prediction with the emulator.} 
\setlength{\tabcolsep}{7pt}
\begin{tabular}{c|cccc}
Isotope & \multicolumn{4}{c}{$e_{\rm max}$} \\
\hline
& 4      & 6      & 8      & 10     \\
\hline
\rule{0mm}{3.5mm}
$^{12}$O     & 293    & 306    & 302    & 85     \\
$^{13}$O     & 294    & 297    & 303    & 30     \\
$^{14}$O     & 863    & 375    & 295    & 70     \\
$^{15}$O     & 285    & 306    & 300    & 78     \\
$^{16}$O    & 292    & 297    & 201    & 43     \\
$^{17}$O     & 301    & 289    & 279    & 50     \\
$^{18}$O     & 291    & 302    & 296    & 50     \\
$^{19}$O     & 287    & 279    & 249    & 60     \\
$^{20}$O     & 305    & 292    & 308    & 68     \\
$^{21}$O     & 266    & 294    & 297    & 65     \\
$^{22}$O     & 280    & 312    & 300    & 74     \\
$^{23}$O     & 304    & 300    & 312    & 69     \\
$^{24}$O     & 302    & 305    & 280    & 72     \\
\hline
\rule{0mm}{3.5mm}
$^{22\!}$Al    & 141    & 58     & 254    & 147    \\
$^{23\!}$Al    & 228    & 263    & 285    & 67     \\
$^{24\!}$Al    & 545    & 541    & 285    & 79     \\
$^{25\!}$Al    & 283    & 268    & 285    & 73     \\
$^{26\!}$Al    & 299    & 288    & 298    & 71     \\
$^{27\!}$Al    & 298    & 282    & 297    & 61     \\
\hline
\rule{0mm}{3.5mm}
$^{28}$Si    & 369    & 130    & 64     & 38     \\
$^{29}$Si    & 358    & 134    & 58     & 37     \\
$^{30}$Si    & 344    & 141    & 46     & 38     \\
$^{31}$Si    & 362    & 139    & 70     & 44     \\
$^{32}$Si    & 365    & 139    & 36     & 36     \\
$^{33}$Si    & 357    & 134    & 70     & 38     \\
$^{34}$Si    & 369    & 151    & 66     & 42     \\
$^{35}$Si    & 350    & 150    & 78     & 28     \\
$^{36}$Si    & 355    & 153    & 75     & 34     \\
$^{37}$Si    & 387    & 150    & 77     & 33     \\
$^{38}$Si    & 377    & 158    & 71     & 34    
\end{tabular}

\label{tab:FrameTraining}
\end{table}

\begin{figure}
    \centering
    \includegraphics[width=\linewidth]{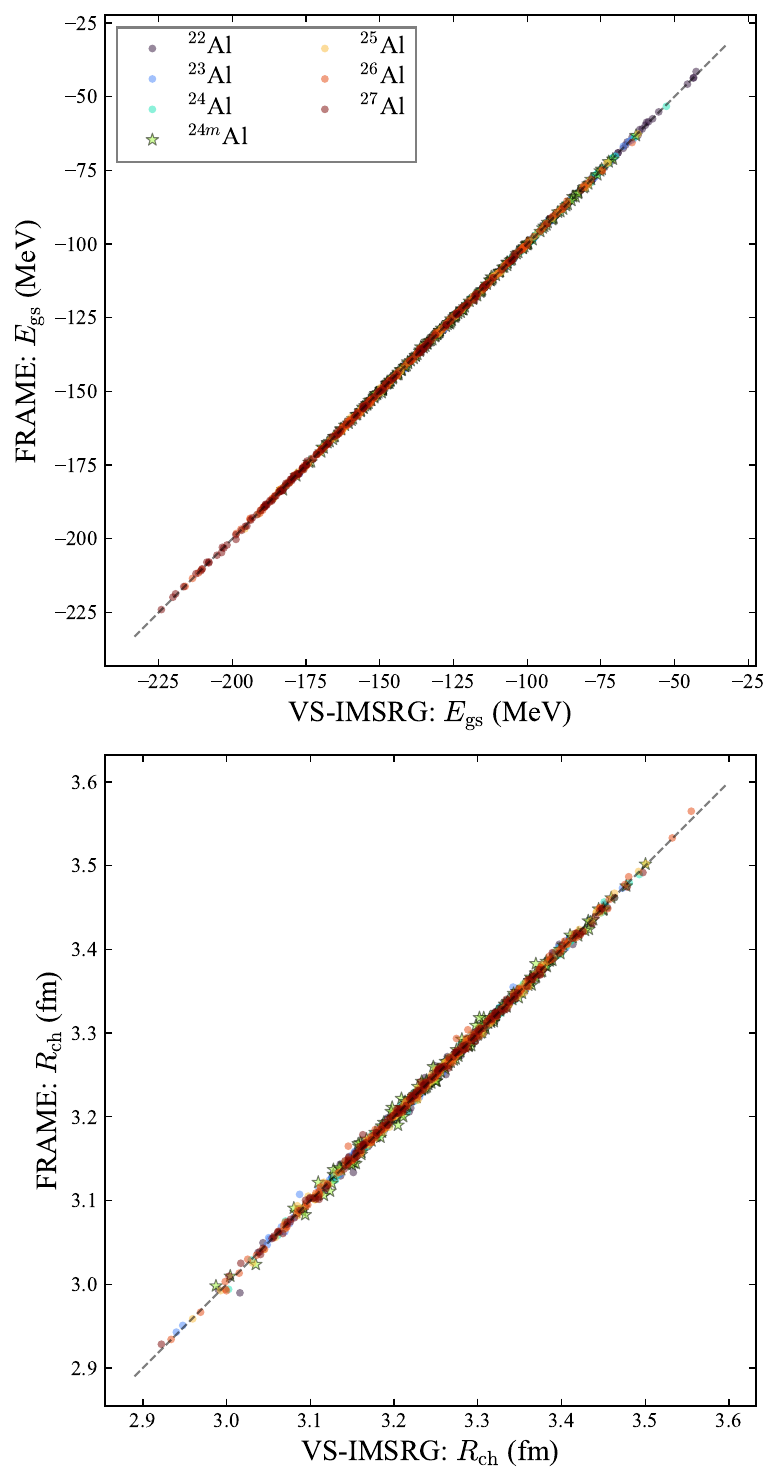}

    \caption{\textbf{Test of emulator.} Calculations of the ground-state energies and charge radii of $^{22\!}$Al--${}^{27\!}$Al using the VS-IMSRG compared to emulated results using the FRAME emulator. For both observables, the emulator reproduces the VS-IMSRG calculations very well, following the line $y=x$ very closely.}
    
    \label{fig:frame_parity}  
\end{figure}

Figure~\ref{fig:frame_parity} shows the comparison between VS-IMSRG predictions and emulated predictions using FRAME for the test dataset for ground-state energies and charge radii. We see that in both cases, the emulator reproduces the exact calculations very well. In particular, we find a root mean-squared error (RMSE) of 0.3 MeV for ground-state energies and 0.003 fm for charge radii. In both cases, the uncertainty of the emulator is negligible compared to the other sources of uncertainties (Hamiltonian uncertainty, many-body method, etc.). We find that including the data from other isotopic chains significantly improves the performance of the emulator, reducing the RMSE from 0.008 fm from the emulator trained only on aluminium data to 0.003 fm with the emulator trained on all data. This highlights the capability of the emulator to learn information across different isotopic chains.

\begin{figure}
    \centering
    \includegraphics[width=\linewidth]{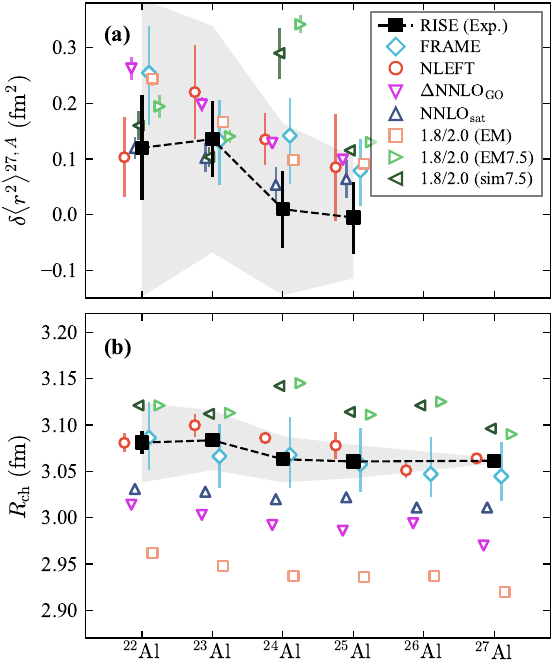}

    \caption{\textbf{Differential mean-squared and absolute nuclear charge radii.}
    Same as Fig.~\ref{fig:diffCharge} but including all interactions considered in this work. See text for more details.}
    \label{fig:figure_rch}  
\end{figure}

The posterior predictive distribution from the over 8000 samples computed using the  FRAME emulator quantifies the emulator and LEC uncertainties.
Our error bars indicate the 68$\%$ credible intervals for the computed distributions. These uncertainties are compared to the NLEFT results and VS-IMSRG results using the 5 interactions described above in Fig.~\ref{fig:figure_rch}.

\begin{table*}[]
\caption{All theoretical values computed for this work. Uncertainties for NLEFT results represent statistical uncertainties from the Monte Carlo, FRAME uncertainties represents total Bayesian evaluation of the Hamiltonian uncertainties and uncertainties from other VS-IMSRG calculations represent model space truncation uncertainties.}
\setlength{\tabcolsep}{7pt}
\begin{tabular}{c|c|ccccccc}
Interaction                                      &                              & $^{22\!}$Al  & $^{23\!}$Al  & $^{24\!}$Al  & $^{24\mathrm{m}\!}$Al & $^{25\!}$Al  & $^{26\!}$Al   & $^{27\!}$Al  \\
\hline\hline
\multirow{3}{*}{NLEFT} & $\langle r_{\rm ch} \rangle$ & 3.081(10) & 3.100(12) & 3.086(5)  & 3.101(15) & 3.078(15) & 3.051(8)   & 3.064 (6)  \\
& $\langle r_{\rm pp} \rangle$ & 2.971(10) & 2.991(12) & 2.979(5)  & 2.994(15) & 2.972(15) & 2.946(8)   & 2.960(6)  \\
& $\langle r_{\rm nn} \rangle$ & 2.730(11) & 2.810(11) & 2.834(5)  &            & 2.901(10) & 2.924(12)  & 2.929(12) \\
\hline\hline
\multirow{3}{*}{FRAME}  & $\langle r_{\rm ch} \rangle$ & 3.085(36) & 3.065(35) & 3.066(39) & 3.112(39) & 3.056(35) & 3.045(33) & 3.042(32) \\
& $\langle r_{\rm pp} \rangle$ & 2.978(38) & 2.961(36) & 2.966(40) & 2.996(41) & 2.957(36) & 2.950(34)  & 2.949(33) \\
\hline\hline
\multirow{3}{*}{$\Delta {\rm NNLO}_{\rm GO}$} & $\langle r_{\rm ch} \rangle$ 
& 3.014(7) & 3.003(4) & 2.992(2)  && 2.986(3) & 2.984(3) & 2.970(3)      \\
& $\langle r_{\rm pp} \rangle$ 
& 2.894(8) & 2.887(4) & 2.878(2) && 2.875(3) & 2.877(3) & 2.864(3)      \\
& $\langle r_{\rm nn} \rangle$ 
& 2.639(1) & 2.707(1) & 2.758(1)  && 2.811(1) & 2.850(1) & 2.880(1)      \\
\hline
\multirow{3}{*}{NNLO$_{\rm sat}$} & $\langle r_{\rm ch} \rangle$ 
& 3.031(1) & 3.028(4) & 3.020(7) && 3.022(8)  & 3.011(9)  & 3.011(9) \\
& $\langle r_{\rm pp} \rangle$ 
& 2.912(1) & 2.913(4) & 2.908(7) && 2.912(9) & 2.904(10) & 2.907(10) \\
& $\langle r_{\rm nn} \rangle$ 
& 2.668(7) & 2.742(9) & 2.791(9) && 2.850(9) & 2.876(8) & 2.921(9)      \\
\hline
\multirow{3}{*}{1.8/2.0~(EM)} & $\langle r_{\rm ch} \rangle$ 
& 2.962(6) & 2.948(2) & 2.937(1) && 2.936(1) & 2.937(1) & 2.920(1)      \\
& $\langle r_{\rm pp} \rangle$ 
& 2.841(7) & 2.830(2) & 2.821(1) && 2.823(1) & 2.828(2) & 2.812(1)      \\
& $\langle r_{\rm nn} \rangle$ 
& 2.574(1) & 2.643(1) & 2.670(1) && 2.755(1) & 2.799(1) & 2.833(2)      \\
\hline
\multirow{3}{*}{1.8/2.0 (EM7.5)} & $\langle r_{\rm ch} \rangle$ 
& 3.121(9) & 3.113(5) & 3.145(7) && 3.111(3) & 3.125(3) & 3.090(1)\\
& $\langle r_{\rm pp} \rangle$ 
& 3.006(9) & 3.001(5) & 3.040(8) && 3.005(3) & 3.023(3) & 2.989(2) \\
& $\langle r_{\rm nn} \rangle$ 
& 2.737(1) & 2.819(1) & 2.886(1) && 2.939(2) & 2.994(3) & 3.005(3)      \\
\hline
\multirow{3}{*}{1.8/2.0 (sim7.5)} & $\langle r_{\rm ch} \rangle$ 
& 3.121(10) & 3.112(6) & 3.142(20) && 3.114(3) & 3.121(3) & 3.096(3) \\
& $\langle r_{\rm pp} \rangle$ 
& 3.007(11) & 3.000(6) & 3.038(22) && 3.009(4) & 3.012(3) & 2.995(2)      \\
& $\langle r_{\rm nn} \rangle$ 
& 2.732(1) & 2.820(1) & 2.886(1) && 2.941(2) & 2.991(3) & 3.011(3)  \\
\hline
\end{tabular}

\end{table*}

\subsection*{Mirror nuclei and proton skin}
\begin{figure}
    \includegraphics[width=1.\linewidth]{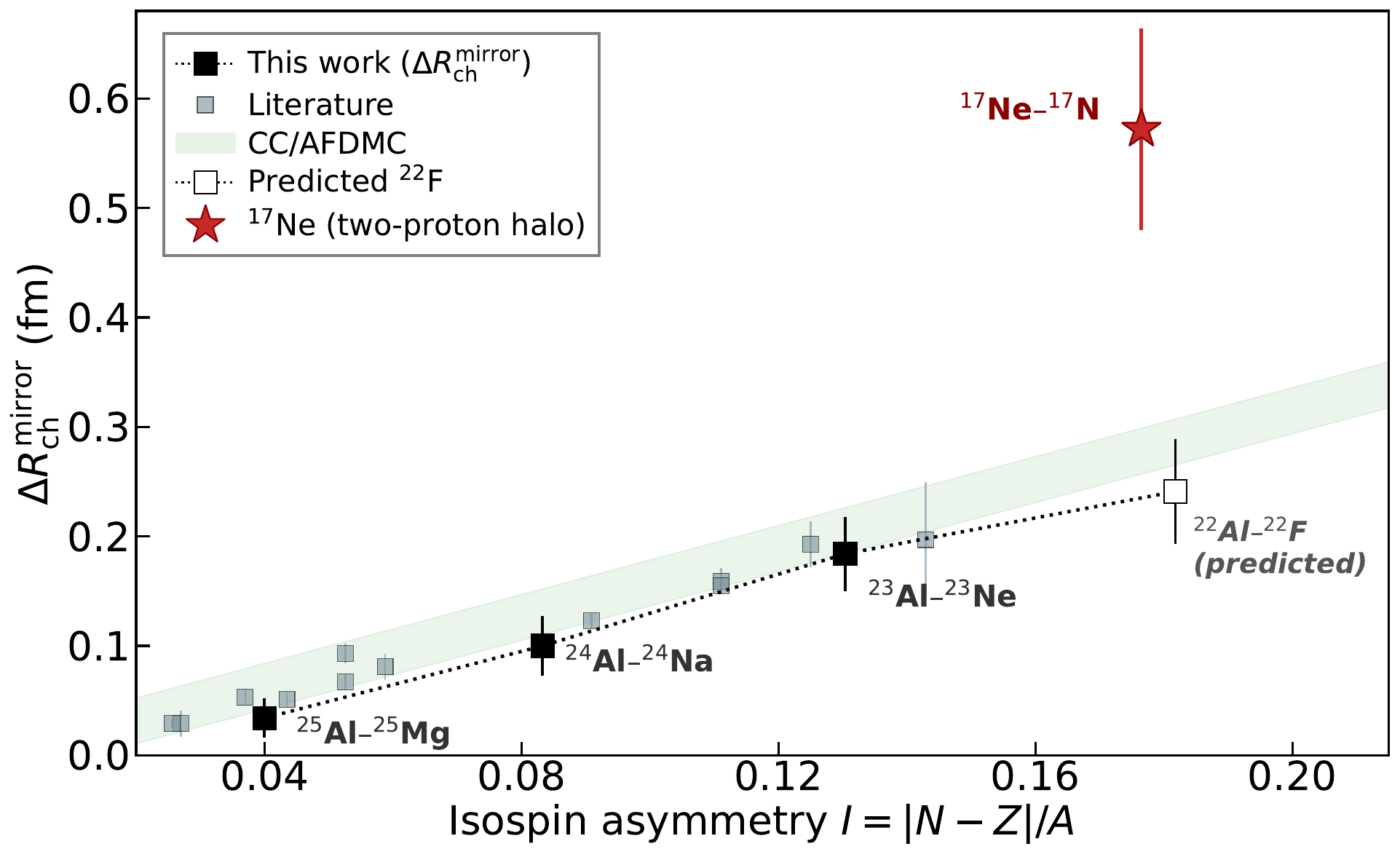}
          \caption{\textbf{Mirror charge-radius differences vs.\ isospin asymmetry.}
Black squares show the results from this work, while grey squares are taken from the literature~\cite{Ohayon2025}. The green band represents the CC/AFDMC prediction~\cite{Novario2023}. The open square denotes the $^{22}$Al--${}^{22}$F mirror charge radii difference, using the present $^{22}$Al result and a predicted $^{22}$F value, from IMSRG calculations, $R_{\mathrm{ch}}$=2.84~fm. The red star indicates the charge radii difference between the measured two-proton halo $^{17}$Ne ($R_{\mathrm{ch}}$=3.042(21)~fm) \cite{Gei08} and its mirror $^{17}$N ($R_{\mathrm{ch}}$=2.47(9)~fm)~\cite{Zhao2024}.}
\label{fig:mirror_shift2}
\end{figure}

Under approximate charge symmetry, the difference in charge radii 
between mirror nuclei is a direct proxy for the proton skin 
 (see Fig.~\ref{fig:proton_skin}). If a proton halo 
were present, the valence proton would extend well beyond the 
neutron core, producing a $\Delta R_{\rm ch}^{\rm mirror}$  significantly larger than 
the systematic trend established for pairs of well-bound mirror 
nuclei~\cite{Ohayon2025,Novario2023}. 
Fig.~\ref{fig:mirror_shift2} tests this expectation by plotting 
the mirror charge radii difference 
$\Delta R_{\rm ch}^{\rm mirror}$ 
as a function of the isospin asymmetry $I = |N-Z|/A$ for all 
mirror pairs with experimentally known charge 
radii~\cite{Ohayon2025} (grey squares), together with our new 
measurements for $^{25\!}$Al--$^{25}$Mg, $^{24\!}$Al--$^{24}$Na, 
and $^{23\!}$Al--$^{23}$Ne (black squares). The green band shows 
the CC/AFDMC prediction 
$\Delta r = 1.574 ~ I \pm 0.021$~fm~\cite{Novario2023}. 
All three aluminium mirror shifts closely follow this band, 
indicating that the observed increase in charge radii toward the 
proton drip line follows the pattern of well-bound nuclei rather 
than reflecting the development of an anomalous proton extension.
To test this hypothesis against a known proton-halo system, we also show the mirror charge-radius difference for $^{17}$Ne--$^{17}$N (red star), using the experimentally measured charge radii $R_{\rm ch}({}^{17}{\rm Ne})=3.042(21)$~fm~\cite{Gei08} and $R_{\rm ch}({}^{17}{\rm N}) = 2.47(9)$~fm~\cite{Zhao2024}. The resulting mirror shift of 0.572(92)~fm is a clear outlier on the mirror trend shown in Fig.~\ref{fig:mirror_shift2}, consistent with the extended proton distribution expected for the two-proton halo in $^{17}$Ne. Recent precision mass measurements of sd-shell dripline nuclei have suggested that mirror energy differences can serve as an indicator of proton-halo formation, suggesting the absence of a halo in the ground states of $^{22,23}$Al but noting that charge radii measurements, reported in this work, were needed to draw definitive conclusions~\cite{Yu2024}. 

\clearpage
\section*{Supplementary Information}

\subsection*{Nuclear charge radii near drip lines}


The purpose of this section is to provide context for the qualitative
charge-radius signature that may be expected from a halo configuration and to
relate this discussion to the present measurements of $^{22,23}$Al. In
particular, $^{22}$Al has a very small proton separation energy and has been
discussed as a possible proton-halo candidate. At the same time, the charge
radii measured in this work do not show a pronounced enhancement of
$^{22}$Al relative to $^{23}$Al, and the mirror charge-radius differences
appear to follow the trend expected for well-bound nuclei. This motivates a
simple estimate of how a possible halo contribution to the charge radius may
depend not only on weak binding, but also on orbital angular momentum and the
Coulomb interaction.

A halo configuration is generally associated with a valence-nucleon wave
function that extends well beyond the nuclear core. Such an extension can be
favored near threshold, but weak binding alone does not necessarily imply a
large contribution to the charge radius. The spatial extent of the valence
particle also depends on the orbital angular momentum and, for protons, on
the Coulomb barrier. The single-particle model discussed below is therefore
not intended as a quantitative description of the many-body charge radius of
$^{22}$Al. Rather, it provides a simple reference point for understanding how
a very weakly bound proton in a higher-$\ell$ configuration can be
consistent with the absence of a clear proton-halo signature in the measured
charge radii.

We assume a finite spherical well potential of depth $V_0$ and radius $R$. The radial wavefunction $u_{l}(r)$ satisfies the Schr\"odinger equation:
\begin{equation}
    -\frac{\hbar^2}{2\mu} \frac{d^2 u_{l}}{dr^2} + \left[ \frac{\hbar^2 l(l+1)}{2\mu r^2} + V(r) \right] u_{l}(r) = -S u_{l}(r),
\end{equation}
where $S$ is the separation energy ($S > 0$ for bound states). The mean-square radius is given by:
\begin{equation}
    \langle r^2 \rangle = \int_0^\infty r^2 |u_{l}(r)|^2 \, dr.
\end{equation}
Near the threshold ($S \to 0$), the behavior of $\langle r^2 \rangle$ is governed by the asymptotic tail of the wavefunction in the classically forbidden exterior region ($r > R$) and the behavior of the asymptotic normalization constant (ANC).

\subsubsection*{Case I: Neutral particle (neutron)}
For a neutron, $V(r) = 0$ for $r > R$. The exterior wavefunction is determined by the centrifugal barrier and is given by the modified spherical Bessel function of the second kind, $k_l(\kappa r)$~\cite{Lu13}:
\begin{equation}
    u_{l}(r) = C_l \sqrt{\frac{2 \kappa r}{\pi}} k_l(\kappa r) \sim C_l e^{-\kappa r} \quad (r \to \infty),
\end{equation}
where $\kappa = \sqrt{2\mu S}/\hbar$ is the inverse decay length and $C_l$ is the normalization constant. Near threshold, matching to the interior wavefunction implies that the square of the ANC scales as $C_l^2 \propto \kappa^{2l+1}$ (up to short-range corrections). This yields the standard halo threshold laws:

\begin{itemize}
    \item \textbf{$s$-wave ($l=0$):}
    \begin{equation}
        \langle r^2 \rangle \sim \frac{1}{\kappa^2} \propto \frac{1}{S}.
    \end{equation}

    \item \textbf{$p$-wave ($l=1$):}
    \begin{equation}
        \langle r^2 \rangle \sim \frac{1}{\kappa} \propto \frac{1}{\sqrt{S}}.
    \end{equation}

    \item \textbf{$d$-wave ($l=2$):} There is no divergence in $\langle r^2 \rangle$ from the tail contribution.  $\langle r^2 \rangle$ approaches a finite constant.
\end{itemize}

\subsubsection*{Case II: Charged particle (proton)}
For a proton, the exterior potential includes the repulsive Coulomb term $V_C(r) = Z e^2/r$. The radial equation for $r > R$ is:
\begin{equation}
    \frac{d^2 u_{l}}{dr^2} - \left[ \kappa^2 + \frac{2\eta \kappa}{r} + \frac{l(l+1)}{r^2} \right] u_{l} = 0,
\end{equation}
where $\eta = Z e^2 \mu / (\hbar^2 \kappa)$ is the Sommerfeld parameter.

Using the WKB approximation in the barrier region,
\begin{equation}
    u_l(r) \sim \exp\left[ - \int\limits_R^r \sqrt{\kappa^2 + \frac{2\eta\kappa}{r'} + \frac{l(l+1)}{r'^2}} \, dr' \right],
\end{equation}
and for $R \ll r \ll \eta/\kappa$, the Coulomb term dominates, giving
\begin{equation}
    u_l(r) \sim \exp\left[ -2\sqrt{2\eta\kappa r} \right].
\end{equation}

This steep decay suppresses the asymptotic tail, and the exterior contribution to the radius scales as $I_{\text{ext}} \propto S^3$, vanishing as $S \to 0$.

\begin{itemize}
    \item \textbf{For $l \ge 1$:} The normalization remains dominated by the interior region, so $\langle r^2 \rangle_p$ remains finite as $S \to 0$.

    \item \textbf{For $l=0$:} Although there is no centrifugal barrier, the Coulomb barrier still suppresses the asymptotic tail strongly enough that the exterior contribution to the radius vanishes as $S \to 0$. The mean-square radius therefore also approaches a finite limit at zero binding. However, this limiting threshold value can be large.
\end{itemize}

For ${}^{22}\mathrm{Al}$, the valence proton is expected to predominantly occupy
the $1d_{5/2}$ orbital ($\ell=2$)~\cite{Campbell2024}. In the simple
single-particle picture discussed above, this implies that the mean-square
radius does not acquire the strong near-threshold enhancement associated with
an $s$-wave halo, and that the spatial extent of the valence proton is
limited by the combined effects of the centrifugal and Coulomb barriers, as
summarized in Table~\ref{tab:scaling}. This provides useful context for the
results discussed in the main text: the measured charge radii of
${}^{22,23}\mathrm{Al}$ are similar within uncertainties, and the mirror
charge-radius differences follow the same systematics as the calculated
proton skins. Thus, while the single-particle estimates presented in this
section are not intended as quantitative predictions for the charge radius of
${}^{22}\mathrm{Al}$, they support the interpretation that the observed radii
are more naturally connected to the gradual evolution of the proton
distribution captured by the many-body calculations than to a strongly
developed halo tail.

\begin{table}[h]
\centering
\caption{Scaling of $\langle r^2 \rangle$ with separation energy $S$ as $S \to 0$. \label{tab:scaling}}
\begin{tabular}{lcc}
\hline \hline
Orbital & Neutron & Proton \\
\hline
$s$-wave & $\propto S^{-1}$ & Finite but can be large \\
$p$-wave & $\propto S^{-1/2}$ & Finite \\
$d$-wave & Finite & Finite \\
\hline \hline
\end{tabular}
\end{table}
\end{document}